\documentclass[aps,preprint,superscriptaddress]{revtex4-1}

\usepackage{amsmath}    
\usepackage{graphicx}   
\usepackage{color}      
\usepackage{hyperref}   
\usepackage{todonotes}
\usepackage{nicefrac}
\usepackage[utf8]{inputenc}

\newcommand{\2}{$2H$-TaSe$_2$}

\begin{document}

\title{Electron Energy-Loss Spectroscopy: A versatile tool for the investigations of plasmonic excitations}
\author{Friedrich Roth}
\email{friedrich.roth@cfel.de}
\affiliation{Center for Free-Electron Laser Science / DESY, Notkestra\ss e 85, D-22607 Hamburg, Germany}
\affiliation{IFW Dresden, P.O. Box 270116, D-01171 Dresden, Germany}
\author{Andreas K\"onig}
\author{J\"org Fink}
\affiliation{IFW Dresden, P.O. Box 270116, D-01171 Dresden, Germany}
\author{Bernd B\"uchner}
\affiliation{IFW Dresden, P.O. Box 270116, D-01171 Dresden, Germany}
\affiliation{Institut f\"ur Festk\"orperphysik, Technische Universit\"at Dresden, D-01062 Dresden, Germany}
\author{Martin Knupfer}
\email{m.knupfer@ifw-dresden.de}
\affiliation{IFW Dresden, P.O. Box 270116, D-01171 Dresden, Germany}
\date{\today}

\begin{abstract}
The inelastic scattering of electrons is one route to study the vibrational and electronic properties of materials. Such experiments, also called electron energy-loss spectroscopy, are particularly useful for the investigation of the collective excitations in metals, the charge carrier plasmons. These plasmons are characterized by a specific dispersion (energy-momentum relationship), which contains information on the sometimes complex nature of the conduction electrons in topical materials. In this review we highlight the improvements of the electron energy-loss spectrometer in the last years, summarize current possibilities with this technique, and give examples where the investigation of the plasmon dispersion allows insight into the interplay of the conduction electrons with other degrees of freedom.
\end{abstract}

\maketitle
\tableofcontents

\section{Introduction}

The exploration of the electronic properties of topical materials represents a major task in condensed matter physics and beyond. Various experimental and theoretical approaches have been developed and steadily improved throughout the last decades. Concerning the experimental techniques, inelastic scattering of particles has become one important branch in order to analyze the electronic excitation spectrum of materials. They often offer the beauty to independently select the scattering angle, and in this way to allow momentum dependent studies of the particular electronic excitation. Inelastic scattering experiments of electron systems require that the probe particles interact with the electrons of the materials under investigation. Therefore two, actually complementary inelastic methods are applied, in particular in condensed matter physics, which are inelastic electron scattering, also called electron energy-loss spectroscopy (EELS) \cite{Raether1979,Schnatterly1979,Lucas1972}, and inelastic light scattering, termed with various acronyms in dependence of the light energy used (Raman scattering, x-ray Raman scattering, inelastic x-ray scattering (IXS)) \cite{Schuelke2007,Kotani2001,Ament2011}.

\par

In this article we describe some recent application of EELS in transmission for the investigation of the collective density excitation of electron gases, the plasmons. In general, EELS is known for many years and has been applied to study the excitations of core electron, valence electrons and vibrations/phonons. This is closely related to the equipment and spectrometers used.

\par

The application of EELS in transmission electron microscopes (TEM) represents one important branch \cite{Egertonbuch,Egerton2009,Abajo2010,Schattschneider1986}. It has been proven to be very useful to detect the chemical and structural composition of the material under scrutiny in combination with a very high lateral resolution. Thereby, core level excitations are predominantly used. Also, using energy filters in combination with microscopy can enhance particular contrasts in the TEM images. The resolution capability of the EELS option in microscopes has been improved during the last years, and also valence band excitation come into the focus of researchers. We note however, that in most cases EELS studies in transmission electron microscopes are not carried out as a function of momentum.

\par

High resolution electron energy-loss spectroscopy (HREELS) carried out in a reflection geometry represents another branch of EELS \cite{Ibach1982,Ibach1993,Richardson1997}. In this case, the electrons are backscattered from a sample surface where they undergo the inelastic scattering process. The primary electron energies used in HREELS are quite low, often in the range of 10\,eV or lower. Low energies are chosen to achieve  very high energy resolution down to or even less than 1\,meV, which renders it possible to study electronic excitations in detail as well as surface vibrations. One of the most important application of HREELS is the investigation of adsorbates on crystal surfaces with many interesting aspects ranging from fundamental surface science to application related questions in ,e.\,g., catalysis. Applying HREELS it should always be kept in mind that one is dealing with a very surface sensitive technique, the scattered electrons do hardly penetrate the material. As a consequence, the excitations studied are surface excitations only, and these can be very different from their bulk counterparts in particular for electronic excitations.

\par

In this contribution, we report on studies using a dedicated spectrometer for EELS in transmission with state-of-the-art resolution parameters and with the option to vary the sample temperature in a rather wide range. In order to elucidate the current status we summarize recent investigations of the behavior of charge carrier plasmons in topical materials, and we discuss the physical insight that is provided by such investigations. The excitation of plasmons, the longitudinal collective excitations of the charge carriers in metals is a specialty of EELS (or IXS) \cite{Raether1979,Schnatterly1979,Fink1989,Schuelke2007}, as it cannot be achieved by light absorption techniques. EELS however is not restricted to the investigation of plasmons, also other excitations such as excitons or inter-band transition can provide valuable information, in particular when probed as a function of momentum transfer \cite{Pichler1998,Neudert1998,Knupfer1999_2,Knupfer2005,Knupfer2002,Schuster2007_penta,Kramberger2008,Roth_2011_b,Crecelius1983,Ritsko1983,Pellegrin1991,FinkLeising1986}.
In addition, we note that the momentum dependence of electronic excitations can be governed by crystal local field effects, a topic that is also studied using EELS in transmission and as a function of momentum transfer \cite{Marinopoulos2002,Waidmann2000,Aryasetiawan1994}.

\section{Instrumentation}

\subsection{Working Principle}
\label{sec:working_principle}

\begin{figure}[h]
\centering
\includegraphics[width=0.8\linewidth]{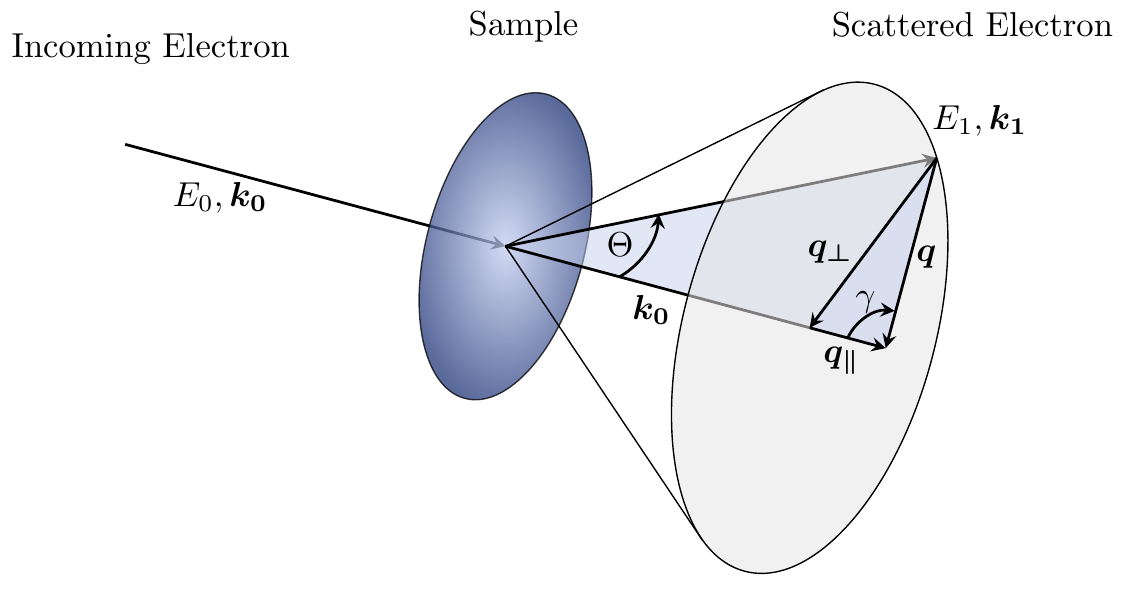}
\caption{The basic scheme for an EELS experiment in transmission. The electrons with an energy $E_0$ and an incoming momentum $\boldsymbol{k_0}$ are focused on the thin sample and scattered under an angle $\theta$. During the scattering event both a momentum transfer $\boldsymbol{q}$ and an energy transfer $\omega$ can occur in the sample. It is possible to perform elastic scattering (Bragg scattering) too by setting the energy loss to zero.} \label{scat_geo}
\end{figure}

The basic scattering geometry of an electron scattering experiment in transmission is shown in Fig.\,\ref{scat_geo}. A beam of rather fast electrons, specified by a momentum $\hbar\boldsymbol{k_0}$, is focused on a thin ($\approx$ 100\,nm) film of the material under investigation. While passing through this sample a fraction of the beam electrons is inelastically scattered by an angle $\theta$ away from the initial direction. This leads to a momentum- ($\hbar \boldsymbol{q}$) and energy-transfer ($\hbar \omega$) given by

\begin{align}
 \hbar \boldsymbol{q} = \hbar \boldsymbol{k_0} - \hbar \boldsymbol{k_1}
\end{align}

\noindent and

\begin{align}
 \hbar \omega = E_0 - E_1 = \frac{\hbar^2(\boldsymbol{k_0^2}-\boldsymbol{k_1^2})}{2m}
\end{align}

\noindent and will as a consequence leave the sample in an excited state, which is characterized by the energy and momentum it acquired from the electrons of the primary beam.

\par

\noindent According to Fig.\,\ref{scat_geo} the momentum transfer $\boldsymbol{q}$ can be decomposed into components parallel and perpendicular to the incoming momentum $\boldsymbol{k_0}$

\begin{align}
 q^2 = q^2_{\parallel} + q^2_{\bot}.
\end{align}

\noindent It is important to realize that the primary energy of the electrons is rather high (the spectrometer described in this contribution works with $E_0$~=~172\,keV). This leads to a large value of $k_0 \sim$ 230\,\AA$^{-1}$. Comparing this to a typical extension of a Brillouin zone inside a solid, which is of the order 1\,\AA$^{-1}$, one arrives at small scattering angles $\theta$ and therefore

\begin{align}
q_{\bot} = k_1 \sin \theta \cong k_1 \theta \cong k_0 \theta.
\end{align}

\noindent Furthermore, the cosine theorem yields (cf. Fig.\,\ref{scat_geo})

\begin{align}
 k^2_1 = k^2_0 + q^2 -2 k_0 q \cos \gamma \cong k^2_0 - 2k_0q \cos \gamma.
\end{align}

\noindent Therefore $q_{\parallel} $ is related to the energy-loss by

\begin{align}
 \hbar \omega \cong \frac{\hbar^2 (k^2_0 - k^2_1)}{2m} \cong \frac{\hbar k_0}{m} \cdot q \cos \gamma = \frac{\hbar k_0}{m} \cdot q_{\parallel}.
\end{align}

\noindent For a typical energy loss around $\hbar \omega \sim$ 10\,eV and at experimentally used scattering angles of $\theta \sim$ 0.25$^\circ$, we obtain the important result

\begin{align}
 \frac{q_\parallel}{q_\bot} = \frac{m \omega}{\hbar k^2_0 \theta} \sim 10^{-3}
\end{align}

\noindent and so the momentum transfer can be regarded as lying completely within the plane which is perpendicular to the incoming electron beam. To obtain information beyond this particular plane one has to rotate the sample with respect to the beam or one has to select other polar directions of $\boldsymbol{k_1}$. Another very important conclusion is that $q \sim q_\bot$ remains valid over a large energy range (up to 100\,eV). This fact is essential for a correct evaluation of the dispersion measurements as well as the Kramers-Kronig analysis.

\par

In addition to the possibility of inelastic scattering processes with non-zero energy-loss it is also possible to perform elastic scattering (Bragg scattering) by setting the energy-loss to zero. This allows the analysis of the crystal structure and is an important tool for investigations of single-crystals where the electronic properties may depend on the direction in the reciprocal lattice.

\subsection{Scattering Theory - The EELS Cross Section}

The essential quantity which is actually measured in the EELS experiment is the doubly differentiated cross section

\begin{align*}
 \frac{\mathrm{d^2} \sigma}{\mathrm{d}\Omega \mathrm{d}\omega},
\end{align*}

\noindent which gives the probability of detecting an electron in a scattering angle element $\mathrm{d}\Omega$ having lost an energy $\mathrm{d}\omega$ compared to its initial energy $E_0$. Quantum mechanically, the scattering process can be described by a transition from an initial state $|n_0, k_0\rangle$ (incident electron plus electrons in the solid in the ground state) to a final state $|n_1, k_1\rangle$ (outgoing electron plus excited electrons in the solid). The interaction of the scattered electrons with the charges in the sample is driven by the Coulomb potential which (in momentum space) is given by

\begin{align*}
 \mathrm{H}_{int} = \frac{e^2}{q^2}.
\end{align*}

\noindent Due to their high kinetic energy, the incident electrons are distinguishable from electrons in the solid and only weak interactions of the electrons with the sample have to be considered. Then the differential cross-section can be written in the Born approximation \cite{Hove1954,Mahan2000}

\begin{align}
  \frac{\mathrm{d^2} \sigma}{\mathrm{d}\Omega \mathrm{d}\omega} = \sum_{n_0,n_1} \left|\langle n_1,k_1 | \mathrm{H}_{int} | n_0,k_0 \rangle \right|^2 \delta (E_{n_0} + E_0 - E_{n_1} - E_1)
\end{align}

\noindent with the initial (final) states of the incoming (outgoing) electrons and the corresponding states for the sample $|n_l\rangle$. The initial and final states can be written as simple products which leads to a new equation for the differential cross-section



\begin{align}
   \frac{\mathrm{d^2} \sigma}{\mathrm{d}\Omega \mathrm{d}\omega} = \left(\frac{\mathrm{d}\sigma}{\mathrm{d}\Omega}\right)_{Ruth} \cdot S(\boldsymbol{q},\omega),
\end{align}

\noindent where

\begin{align*}
 \left(\frac{\mathrm{d}\sigma}{\mathrm{d}\Omega}\right)_{Ruth} = \frac{4}{a^2_0 q^4}
\end{align*}

\noindent is the elastic Rutherford cross-section with $a_0$ being the Bohr radius. The dynamic structure factor $S(\boldsymbol{q},\omega)$ is defined by

\begin{align}
S(\boldsymbol{q},\omega) = \frac{1}{N} \sum_{n_0,n_1} p_{n_0} \left| \langle n_1 | \sum_{j} \mathrm{e}^{\mathrm{i} q r_j} | n_0 \rangle \right|^2 \delta (E_{n_0} - E_{n_1} + \omega).
\end{align}

\noindent The factor $\nicefrac{1}{N}$ has been included because the cross-section is defined per electron. This close relationship between the dynamic structure factor and the density-density correlation function was first derived by Van Hove \cite{Hove1954}.







With the help of the Kubo formalism of linear-response theory \cite{Kubo1957,HANKE1978} and the fluctuation-dissipation theorem \cite{Nyquist1928,Callen1951} one may establish a relation between the dynamic structure factor $S(\boldsymbol{q},\omega)$ and the dielectric function $\epsilon(\boldsymbol{q},\omega)$. This is an example of a very general principle in statistical physics, namely the fluctuation-dissipation theorem which always connects some sort of correlation function (the density-density correlation in this case) with a response function ($\epsilon(\boldsymbol{q},\omega)$)

\begin{align}
 S(\boldsymbol{q},\omega) = \frac{q^2}{4\pi e^2} \frac{1}{1-\mathrm{e}^{-\beta \omega}} \operatorname{Im} \left(-\frac{1}{\epsilon(\boldsymbol{q},\omega)}\right),
\end{align}

\noindent with $\beta = \nicefrac{1}{kT}$ ($k$ is the Boltzmann's constant). For typical electronic excitation energies $\omega \gg \nicefrac{1}{T}$ and neglecting the Bose factor in the previous relation we obtain the final result

\begin{align}
  \frac{\mathrm{d^2} \sigma}{\mathrm{d}\Omega \mathrm{d}\omega} = \frac{\mathrm{const.}}{q^2} \cdot \underbrace{\operatorname{Im} \left(-\frac{1}{\epsilon(\boldsymbol{q},\omega)}\right)}_{\mathrm{Loss~Function}},
\end{align}

\noindent which relates the intensity measured in an EELS experiment to the dielectric function which provides access to the electronic structure of a sample under investigation.

\par

As the dielectric function---and also the loss function---are so-called response functions they reveal several very useful properties. First, from the fact that the response of the system is causal, one obtains the Kramers-Kronig (KK) relations

\begin{align}
 \operatorname{Re} \left( \frac{1}{\epsilon(\boldsymbol{q},\omega)} \right) -1 &= \frac{1}{\pi} \mathcal{P} \int \frac{\mathrm{d} \omega'}{\omega' - \omega} \left[\operatorname{Im} \left( \frac{1}{\epsilon(\boldsymbol{q},\omega ')} \right) \right]\\
\operatorname{Im} \left( \frac{1}{\epsilon(\boldsymbol{q},\omega)} \right) &= -\frac{1}{\pi} \mathcal{P} \int \frac{\mathrm{d} \omega'}{\omega' - \omega} \left[ 1-\operatorname{Re} \left( \frac{1}{\epsilon(\boldsymbol{q},\omega ')} \right) \right],
\end{align}

which are an essential tool to deduce the complete dielectric function from the signal measured in the EELS experiment ($\mathcal{P}$ denotes the Cauchy principal value or principal part of the integral). With this it is possible to derive, in principle, all optical constants like, e.\,g., the optical conductivity, the refractive index etc. \cite{Dressel2002}. Besides the KK equations there exist more relations, so called sum rules, that are important for the evaluation of the data as well as for the calibration of the loss function and the consistency check of our KK analysis. One of these sum rules relates the loss function and the dielectric function to the density $N$, of all valence electrons:

\begin{align} \label{eq:sum_rule}
 \int\limits_{0}^{\infty} \mathrm{d}\omega\, \omega \operatorname{Im}\left(-\frac{1}{\epsilon(\boldsymbol{q},\omega)}\right) = \int\limits_{0}^{\infty} \mathrm{d}d\omega\, \omega \epsilon_2 = \omega_p^2 \cdot \frac{\pi}{2} \propto N
\end{align}

where the plasma frequency is given by

\begin{align}
\label{Plasma_Freq}
\omega_p = \sqrt{\frac{4\pi N e^2}{m_e}}.
\end{align}

Thus, the strengths of possible transitions are not independent from each other but are balanced in a way that enhancing the weight in a particular energy range of the spectrum by, e.\,g., the appearance of a phase transition will reduce the intensity in another energy window to keep the above given integral at a constant value. In practice, calculations as well as experiments are of course always restricted to a finite energy window and one may evaluate partial sum-rules according to

\begin{align*}
 \int\limits_{\omega_0}^{\omega_1} \mathrm{d}\omega \omega \operatorname{Im}\left(-\frac{1}{\epsilon(\boldsymbol{q},\omega)}\right) = \omega_p^2 \cdot \frac{\pi}{2} (N_{eff}/N),
\end{align*}

\noindent that provide access to an effective number of charge carriers $N_{eff}$ contributing to a particular type of excitation within a given energy range.

Finally, for metallic systems a further sum rule can be employed \cite{Mahan2000}, which may allow an additional check of the data analysis:

\begin{align} \label{eq:sum_rule2}
 \int\limits_{0}^{\infty} \mathrm{d}\omega \frac{\operatorname{Im}\left(-\frac{1}{\epsilon(\boldsymbol{q},\omega)}\right)}{\omega} = \frac{\pi}{2}.
\end{align}

\noindent Moreover, measurements of the resistivity represent another way to observe how several degrees of freedom influence the conduction electrons. Interestingly, there exist a simple relation between the resistivity (reciprocal value of the (DC)-conductivity $\sigma$) and the plasmon properties:

\begin{align} 
\sigma =  \omega_p^2 \cdot \epsilon_0 \cdot \tau,
 \end{align}

where $\epsilon_0$ is the background dielectric constant and $\tau$ displays the scattering rate (correlated with the width of the plasmon) \cite{Gross2012}.

\subsection{Experimental Details}

\subsubsection{The Spectrometer}

\begin{figure}[h]
\centering
\includegraphics[width=0.7\linewidth]{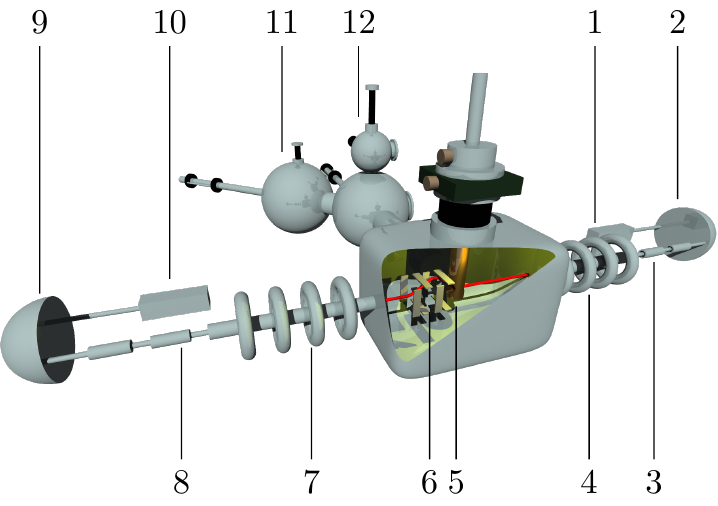}
\caption{Schematic drawing of the electron energy-loss spectrometer. (1\,-\,Source, 2\,-\,Monochromator, 3\,-\,Zoom lenses,  4\,-\,Accelerator, 5\,-\,Sample holder with cryostate, 6\,-\,Deflection plates, 7\,-\,Decelerator, 8\,-\,Zoom lenses, 9\,-\,Analyser, 10\,-\,Multiplier, 11\,-\,Preparation chamber, 12\,-\,Sample magazine and Fast entry)} \label{spectromter}
\end{figure}

In Fig.\,\ref{spectromter} we show the main parts of the transmission electron energy-loss spectrometer as operated in the IFW Dresden. The spectrometer parts will be described briefly below, for a more detailed description we refer the reader to previous articles \cite{Fink1989,Fuggle1992}.

Two major advancements have been achieved recently which enhance the capabilities of the spectrometer substantially. First, the electron optics has been recalculated (for details about such calculations see \cite{Fink1989,Fink1980}) and changed accordingly in order to improve the energy/momentum resolution (see below) with even higher electron flux. Second, a cryostat with a special sample stage has been introduces to allow for measurements as a function of temperature.

The entire spectrometer is placed in an ultra-high vacuum (UHV) environment. The electron source consists of a (osmium coated) tungsten cathode followed by a lens system that focuses the electron beam to the entrance slit of the monochromator. Afterwards, the electron beam is guided by so called zoom lenses to the accelerator that increases the energy of the electrons to a value of $E_0$~=~172\,keV. After passing through the sample, momentum selection of the scattered particles is achieved by two pairs of deflection plates, arranged horizontally and vertically, forcing the scattered electrons back to the optical axis. Furthermore, the energy loss is determined by adding an additional voltage to the acceleration voltage. After being decelerated and passing through further electrostatic lenses, the electrons reach the analyzer and finally the detector, where a photo multiplier produces the signal transferred to the computer. Both, monochromator and analyzer are hemispherical systems, as frequently found in photoelectron spectroscopy \cite{Damascelli2003}.

The sample position with respect to the primary electron beam can be chosen via a UHV manipulator. With the manipulator it is possible to change the sample position within the plane perpendicular to the beam axis as well as to change the horizontal angle between sample plane and beam direction. Furthermore, the manipulator is now equipped with a helium flow-cryostat and a temperature controller, allowing measurements in a wide temperature regime of $T~\approx~20...400$\,K. Due to the transmission setup of the spectrometer and to keep the samples transferable within the UHV set-up, a direct contact of the sample holder to the cryostat can only be realized by special, adjustable metal clamps. For this reason the reproducibility of fixing the temperature only is around $5$\,K. Equipping the spectrometer with a cryostat is of great importance since it on the one hand increases the lifetime of the samples, especially those of organic compounds and on the other hand decreases thermal peak broadening. Moreover, it gives access to temperature dependent physical effects such as phase transitions or freezing out of specific excitations.

\begin{figure}[h]
\centering
\includegraphics[width=0.45\linewidth]{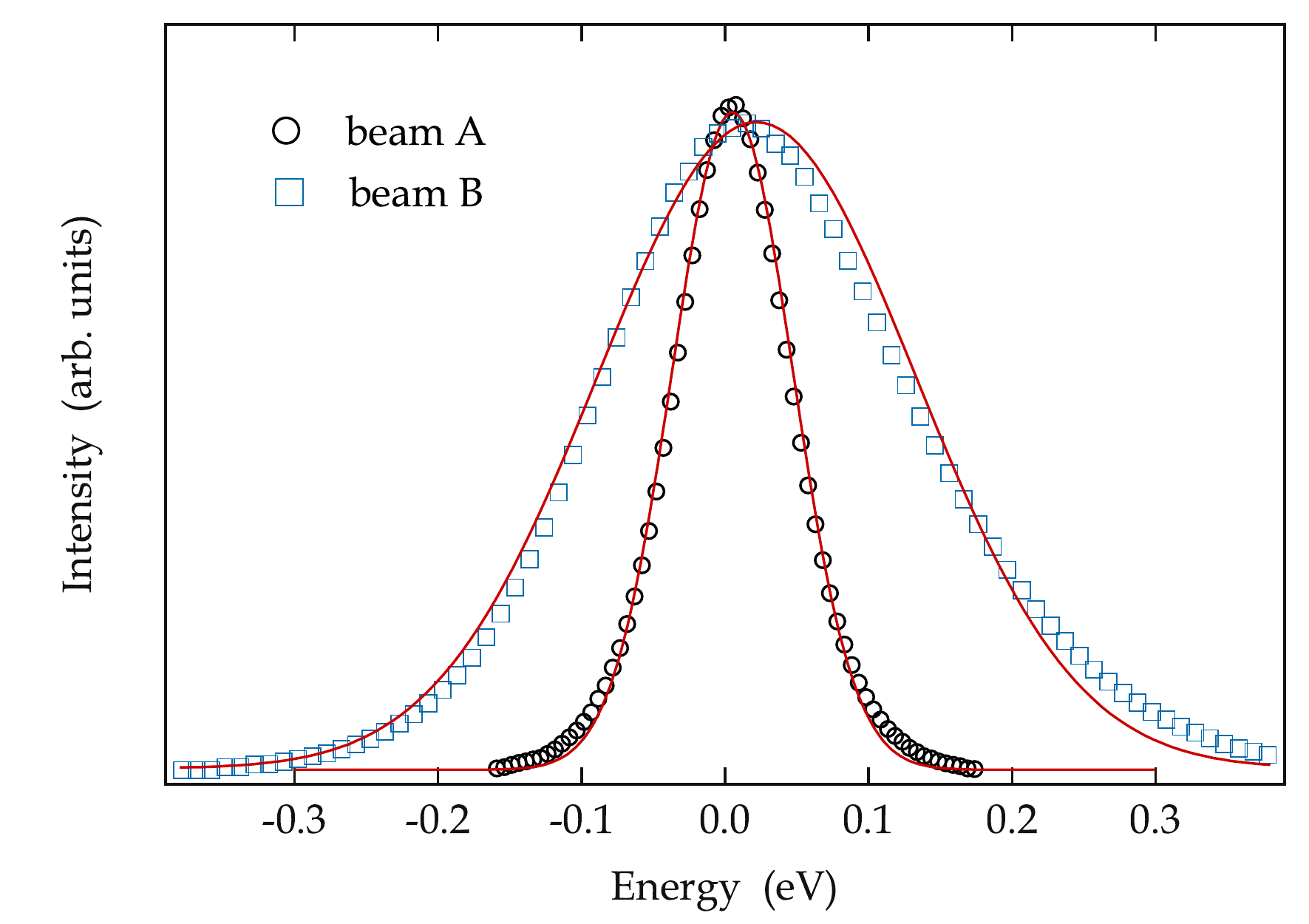}
\includegraphics[width=0.45\linewidth]{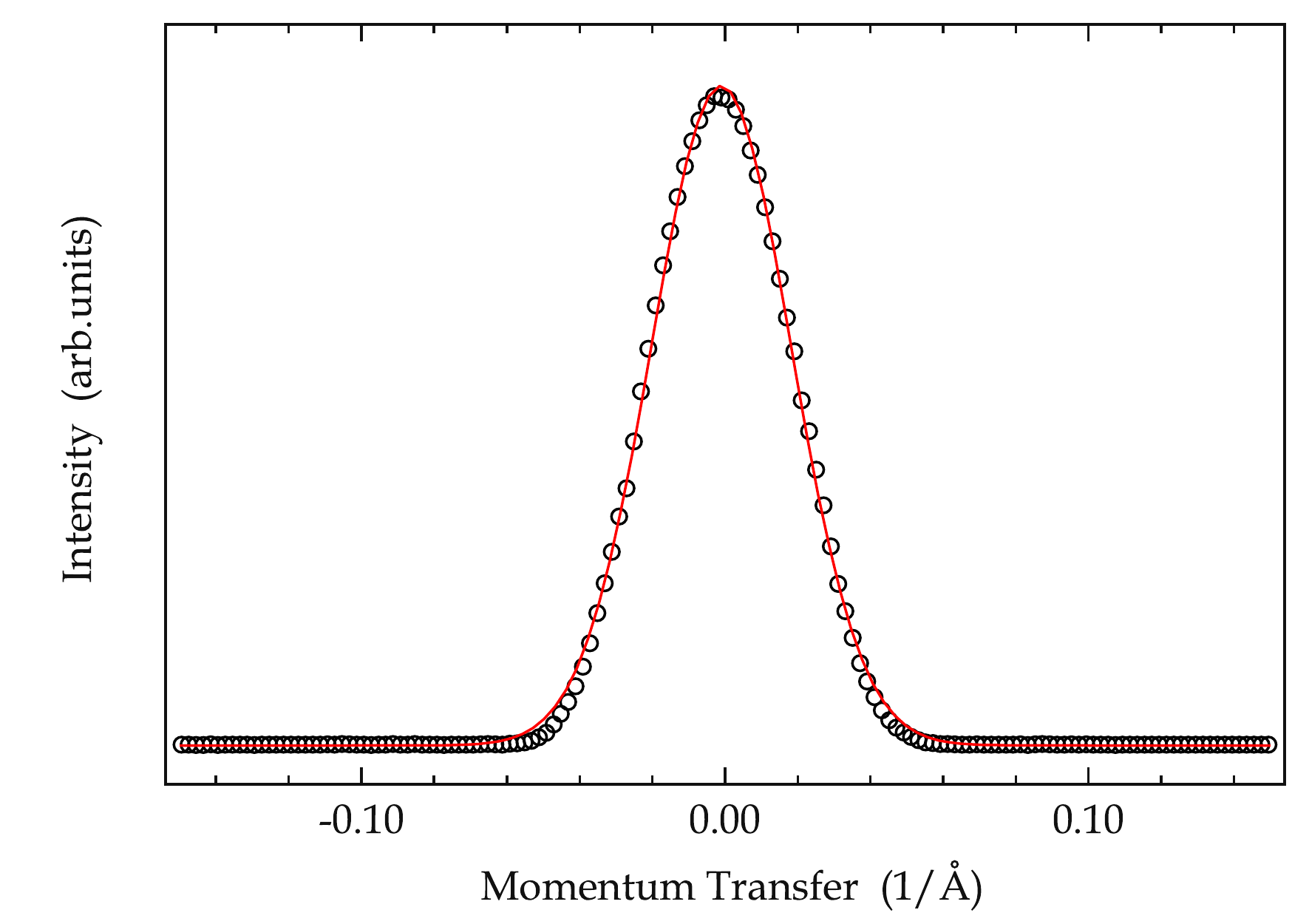}
\caption{The energy and momentum resolution curves for the primary electron beam (without sample). The data (black circles - beam A used for investigation of valence band excitations; blue diamonds - beam B used for investigations of core-level excitations) fitted to Gaussians (red solid line). The obtained FWHM values are $\Delta E_{\mathrm{beam~A}} \approx$ 85\,meV and accordingly $\Delta E_{\mathrm{beam~B}} \approx$ 200\,meV as well as $\Delta q \approx$ 0.03\,\AA$^{-1}$ (notice that the momentum resolution for both beams is equal).} \label{resolution}
\end{figure}

For an optimal investigation of the distinct types of electronic excitations---valence- as well as core-level excitations---different beam characteristics are possible. The settings required for a certain beam are stored in files containing all necessary adjustments for the power supplies, deflection plates and so on. Nevertheless, each beam path is tuned by an automatic procedure before a new sample is loaded for best possible performance. The characteristic data (energy and momentum resolution) for the two electron beams most often used are plotted in Fig.\,\ref{resolution}. They are fitted with Gaussians according to

\begin{align*}
 I(q) &\propto \frac{1}{\sigma_q} \cdot \exp\left[-\left(\frac{q}{\sigma_q}\right)^2\right]\\
 I(E) &\propto \frac{1}{\sigma_E} \cdot \exp\left[-\left(\frac{E}{\sigma_E}\right)^2\right],\\
\end{align*}

\noindent with the widths $\sigma_q$ and $\sigma_E$. The energy and momentum resolution (full width at half maximum, FWHM) of 85\,meV and 0.03\,\AA$^{-1}$ are applied for valence band excitations and 200\,meV and 0.03\,\AA$^{-1}$ for core-level excitations, respectively (cf.~Fig.\,\ref{resolution}). One has to keep in mind that the energy resolution is controlled by the pass energy of the monochromator and the analyser, while the momentum resolution is set by the optical properties of the zoom lenses. Moreover, the ripple of the high voltage power supply has no influence on the energy resolution as long as the time of flight of the electrons is much faster than the inverse frequency of the ripple.

\begin{figure*}
\centering
\includegraphics[width=0.45\linewidth]{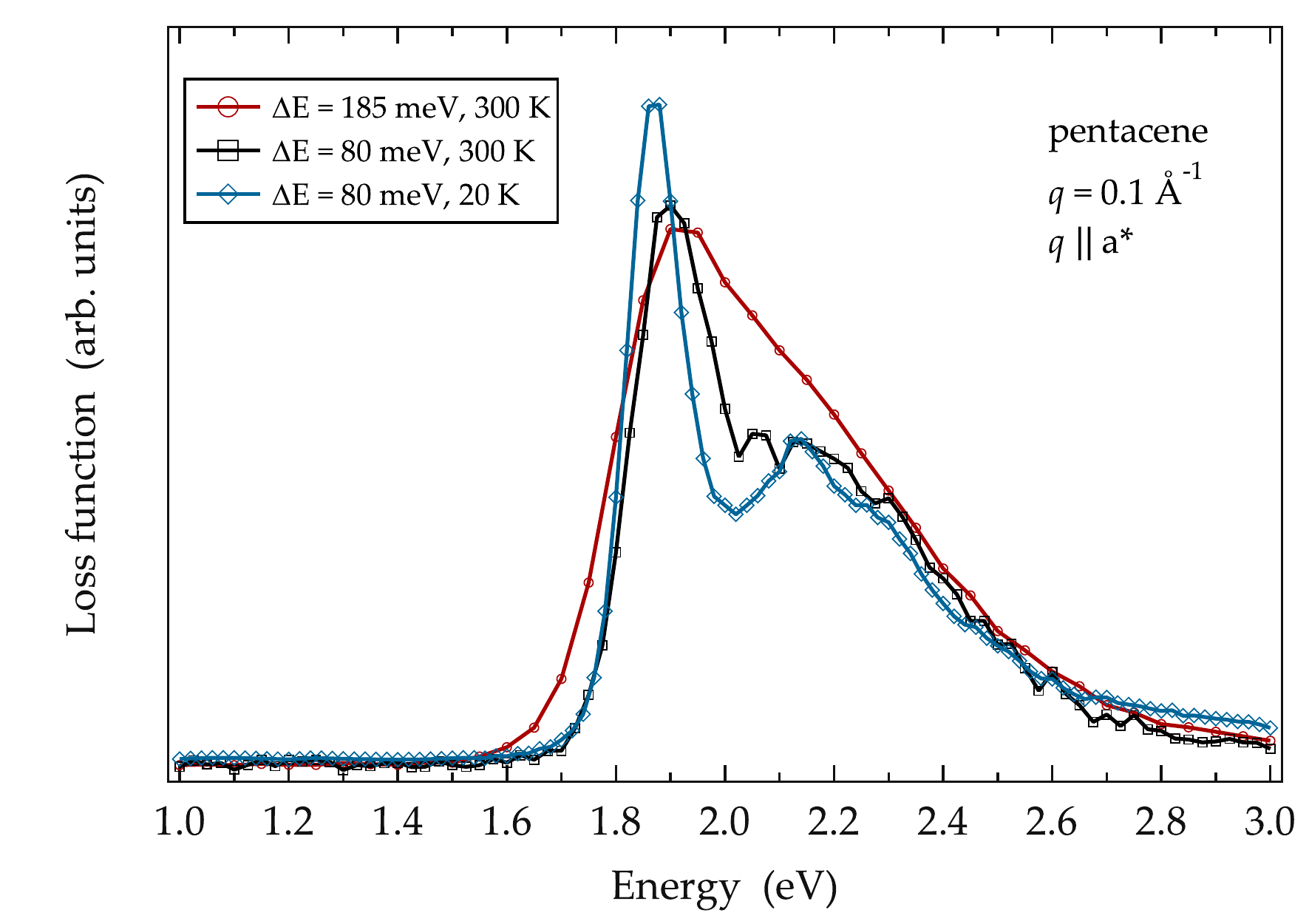}
\includegraphics[width=0.45\linewidth]{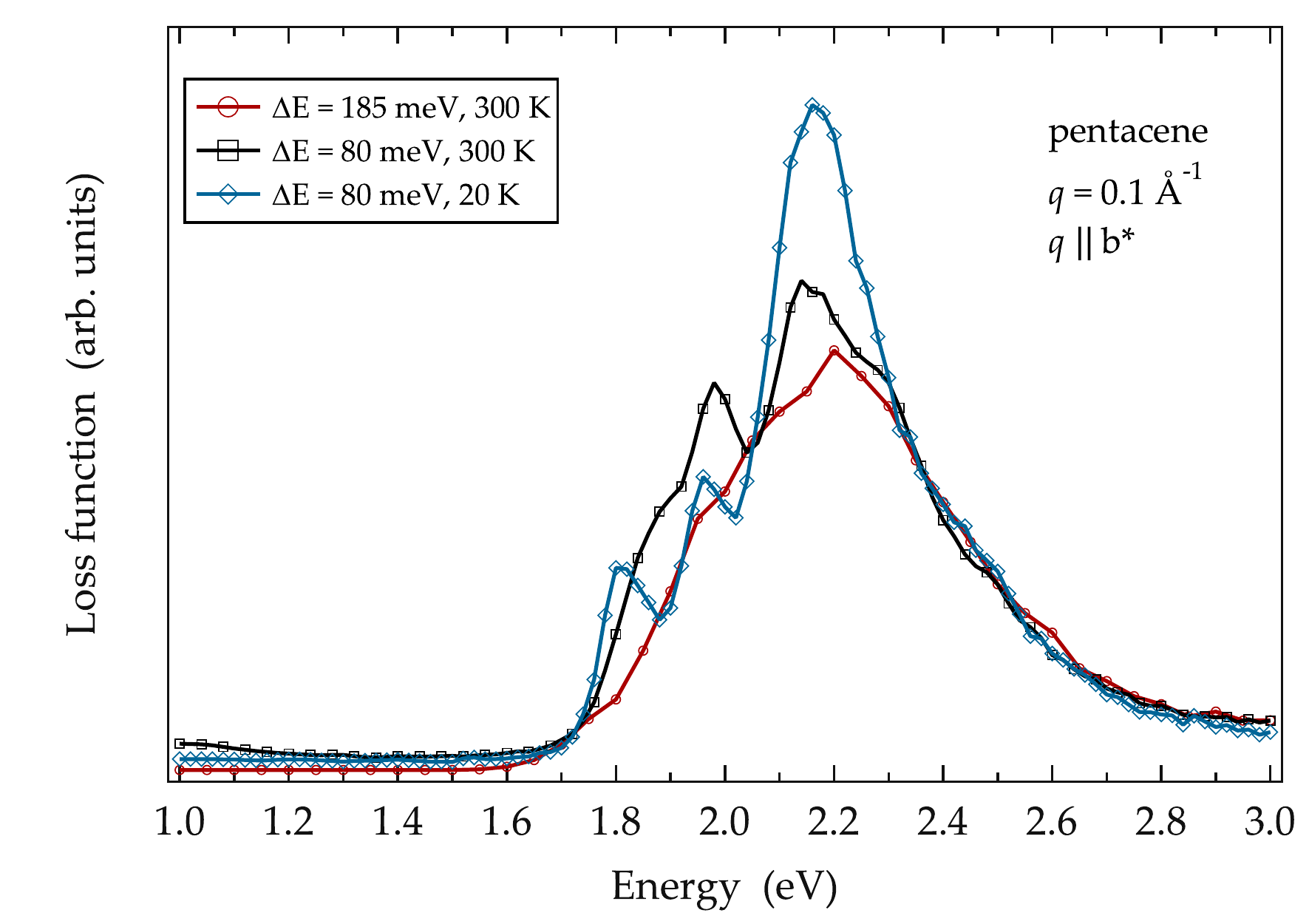}
\caption{Comparison of the loss function of pentacene single crystals for $q \parallel a^*$ (left panel) and $q \parallel b^*$ (right panel) as a function of the energy resolution of the used electron beam (red circles) as well as the temperature (black squares - room temperature, blue diamonds - 20\,K).} \label{pentacene}
\end{figure*}

In Fig.\,\ref{pentacene} we show one impressive example how improving the energy resolution on the one hand and the new possibility to cool the samples down to 20\,K on the other hand enhance the information depth that can be achieved, and associated with give the opportunity for entirely new and sometimes unexpected insights. The spectra, representing singlet excitons in pentacene for small momentum transfer along the fundamental reciprocal lattice vectors, which were measured some years ago with an energy resolution of about 185\,meV at 300\,K show now visible structure and are represented by a broad asymmetric feature \cite{Schuster2007_penta}. The situation changes significantly by measuring the equivalent spectra with a better energy resolution of 80\,meV and further by cooling the samples down to 20\,K. As shown in Fig.\,\ref{pentacene} one now can identify a fine structure in the energy range of the former broad excitation feature (for further reading see Ref.\,\cite{Roth_penta}). Thus, the much higher resolution allows the identification of several excitonic/ vibronic features, the exact nature of which is not completely understood yet.

\subsubsection{Sample Preparation}

One of the major challenges in the application of EELS in transmission in solid-state physics and material science is the preparation of the samples. The mean free path of electrons in solids is limited due to a number of interactions, predominantly plasmon excitations \cite{Egertonbuch,Egerton1987,Wang2010,Pozsgai2007,Zhang2012}.

Therefore samples with a thickness of only about 100\,nm are required and form the main criterion whether or not a system can be investigated by EELS.
Furthermore, the electron beam in the spectrometer described here has a focal size of the order of 0.5 mm$^2$, i.\,e. the samples also must have a similar lateral extension. There exist several possibilities to obtain the required thickness of the films depending on the microscopic structure of the actual material. On the one hand large thin films of e.g. organic compounds can be produced by thermal evaporation under high vacuum onto single crystalline substrates (e.\,g. KBr) kept at room temperature in a separate vacuum chamber. With that technique it can even be possible to influence the crystal growth and the crystal orientations by controlling the deposition rate and evaporation temperature.

\begin{figure}[h]
\centering
\includegraphics[width=0.5\linewidth]{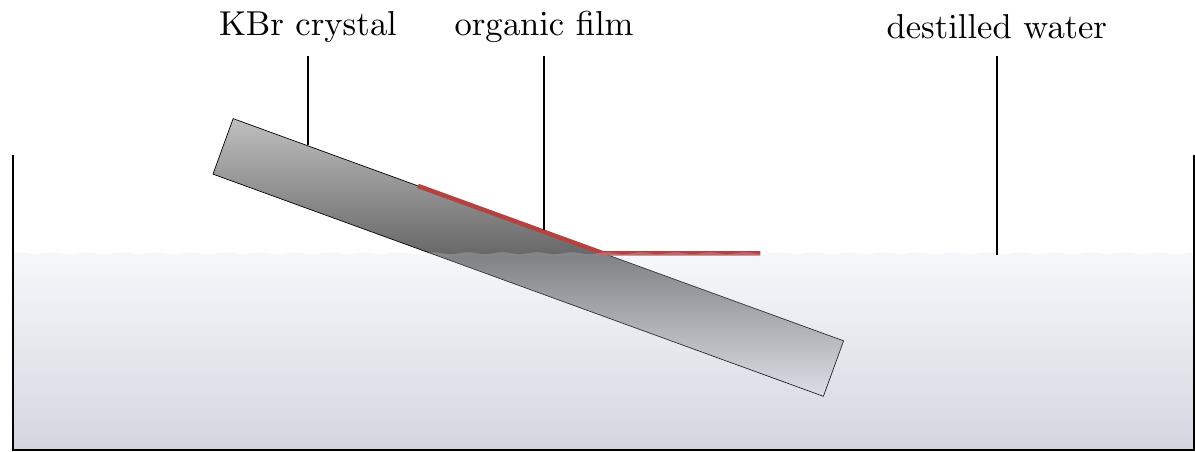}\qquad
\includegraphics[width=0.3\linewidth]{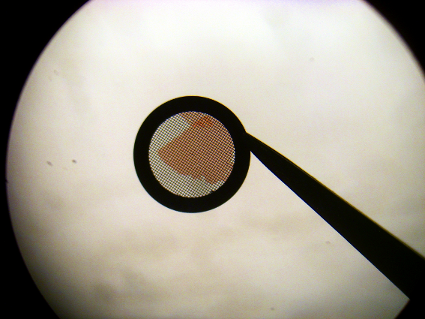}
\caption{Left panel: Simplified view of the preparation process of the thin films for the investigation using the EELS spectrometer. The evaporated organic film is floated off in distilled water until it is separated from the substrate. Afterwards, small pieces of the film can be mounted onto standard TEM grids and transferred into the spectrometer. Right panel:  A typical snapshot of a thin film placed on a standard TEM grid (diameter $\approx$ 5\,mm) intended for usage in EELS measurements.} \label{preparation}
\end{figure}

 Subsequent to the evaporation the films have to be separated from the KBr substrate. For that purpose the films are floated off in distilled water until the thin film is detached from the substrate and floats at the water surface due to the surface tension (cf. left panel of Fig.\,\ref{preparation}). Afterwards, small pieces of the film are mounted onto standard electron microscopy grids (see Fig.\,\ref{preparation} (right panel)), incorporated into an EELS sample holder, and transferred into the spectrometer.

The preparation of thin films starting with single crystals obviously requires a different approach. With the help of an ultramicrotome ---a special device allowing precise cuts with the help of a diamond knife--- it is possible to cut thin slices from a macroscopic single crystal. Subsequently, the resulting films have again to be transferred onto a standard electron microscopy grid. This method however can suffer from sample hardness or brittleness which limits its applicability. Alternatively, one can also try to repeatedly cleave the respective material using an adhesive tape until films of about 100 nm thickness have been achieved. Later, the adhesive tape must be dissolved in appropriate organic solvents. A well known example would be the preparation of thin graphite samples \cite{Roth_HOPG}, where it is even possible to thin the crystal down to a monolayer of graphene (also called exfoliation of graphene \cite{Geim2009}).

\begin{figure}[h]
\centering
\includegraphics[width=0.55\linewidth]{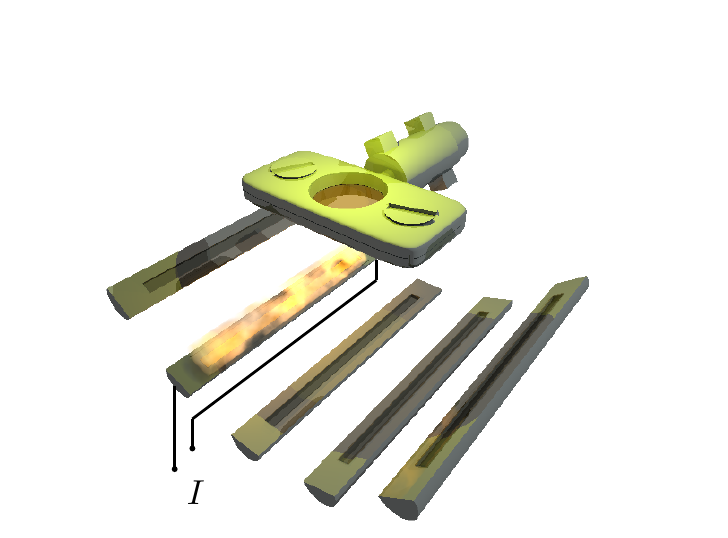}
\includegraphics[width=0.35\linewidth]{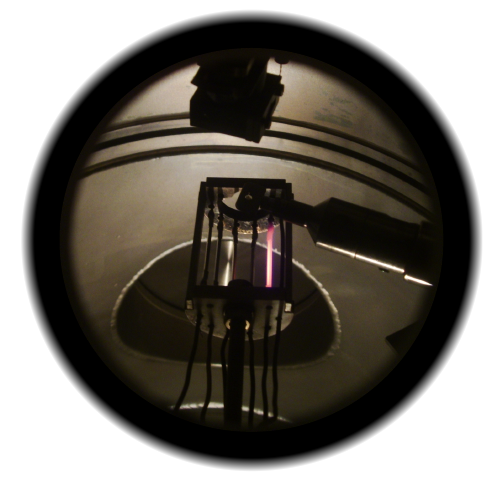}
\caption{Left panel: Schematic drawing of a sample holder and the \emph{in~situ} doping with alkali metals evaporated from commercial getter sources. Right panel: Snapshot of the construction in the vacuum chamber. At the bottom of the picture one can see the holder for the getter sources (notice that the fifth getter is in an operating state). Furthermore, one can identify the transfer rod with a sample holder on top as well as the oven in the upper part of the picture.} \label{getter}
\end{figure}

In addition, the UHV set-up of our EELS spectrometer including a preparation chamber also enables to further modify the samples by for instance heating to temperatures up to about 800\,K in a dedicated sample furnace and under inert conditions. Another option is the evaporation of further materials onto the samples, e.\,g., it is possible to intercalate the particular samples with alkali metals in the preparation chamber. For this, the sample is exposed to an alkali metal atmosphere from commercially available alkali metal getter sources (distance between getter source and sample is about 30\,mm)(cf.~Fig.~\ref{getter}).

\section{Selected Results: The investigation of plasmons in topical materials}

The most prominent excitations of the electron gas for small momentum transfers are collective oscillations. Quantum-mechanically these excitations are described as quasi particles and are called \emph{plasmons}. In consideration of the conduction electrons as a quasi-free electron gas, the density of this electron gas is unstable towards an external electrostatic perturbation at a characteristic frequency, i.\,e., it will oscillate at this characteristic frequency $\omega_p$ (see Eq.\,\ref{Plasma_Freq}) like a harmonic oscillator provided that its wavelength is longer than a characteristic cutoff $\nicefrac{1}{q_c}$, which is determined by the decay of the density oscillations into intra-band excitations ($n$ is the density of the electron gas, $e$ denotes the elementary charge, $\epsilon_\infty$ the background dielectric screening due to interband excitations, and $m^*$ the (optical) effective mass of the charge carriers).  The quanta of these collective density fluctuations were first coined plasmons by David Pines in Ref.\,\cite{Pines1956}.

A feature of special importance related to plasmons is the momentum dependence of the plasma frequency, i.\,e., the plasmon dispersion which can be derived from the
Lindhard function under some assumptions \cite{Raimes1957,LINDHARD1954,Nolting2009}. As a result, the plasmon dispersion can be expanded into a Taylor series at $\boldsymbol{q}$ = 0 given by

\begin{align}
 \omega(\pmb{q}) = \omega_p +\alpha \pmb{q}^2 + \mathcal{O} (\pmb{q}^4); \quad \alpha \propto v_F^2,
\label{eqn:omega_q}
\end{align}

where $v_F$ is the mean Fermi velocity. This is the generic behavior of the collective plasmon modes in the random phase approximation (RPA) and there are numerous examples in the literature, that this shape of plasmon dispersion is realized for different simple metals, such as Li and Na~\cite{Gibbons1976} as well as for more complex structures, such as the planar cuprates, e.\,g., Bi$_2$Sr$_2$CaCu$_2$O$_8$ \cite{Nucker1989}. However, there are also examples for anomalous dispersions, varying from linear to even negative slope. They contain important physical information on the modification of the electron gas by other degrees of freedom in the material. This will be further emphasized in the following sections, which demonstrate that EELS is a versatile spectroscopic method for the study of condensed matter and has become a standard method for investigating the collective excitations of electrons.

\subsection{Peculiar plasmon dispersion in potassium intercalated picene}

In the next section we present measurements on a recently reported organic superconductor, potassium doped picene \cite{Mitsuhashi2010}. Picene (C$_{22}$H$_{14}$) is a molecule that consists of five benzene rings arranged in a zigzag like manner and forms a herringbone, monoclinic crystal structure, similar to other aromatic molecules \cite{De1985}.

The addition of potassium results in a charge transfer of the outer $s$ electrons to the picene molecules, i.\,e., a doping of the molecular crystal. For such crystals superconductivity has been reported with a transition temperature up to 18\,K  \cite{Mitsuhashi2010}, which is rather high as compared to many other doped molecular crystals or charge transfer salts. The only molecular materials that shows superconductivity at higher temperatures are the fullerides, the discovery of which represented a breakthrough in the field of superconductivity \cite{Hebard1991,Gunnarsson2004}.

\begin{figure*}[h]
\centering
\includegraphics[width=0.48\linewidth]{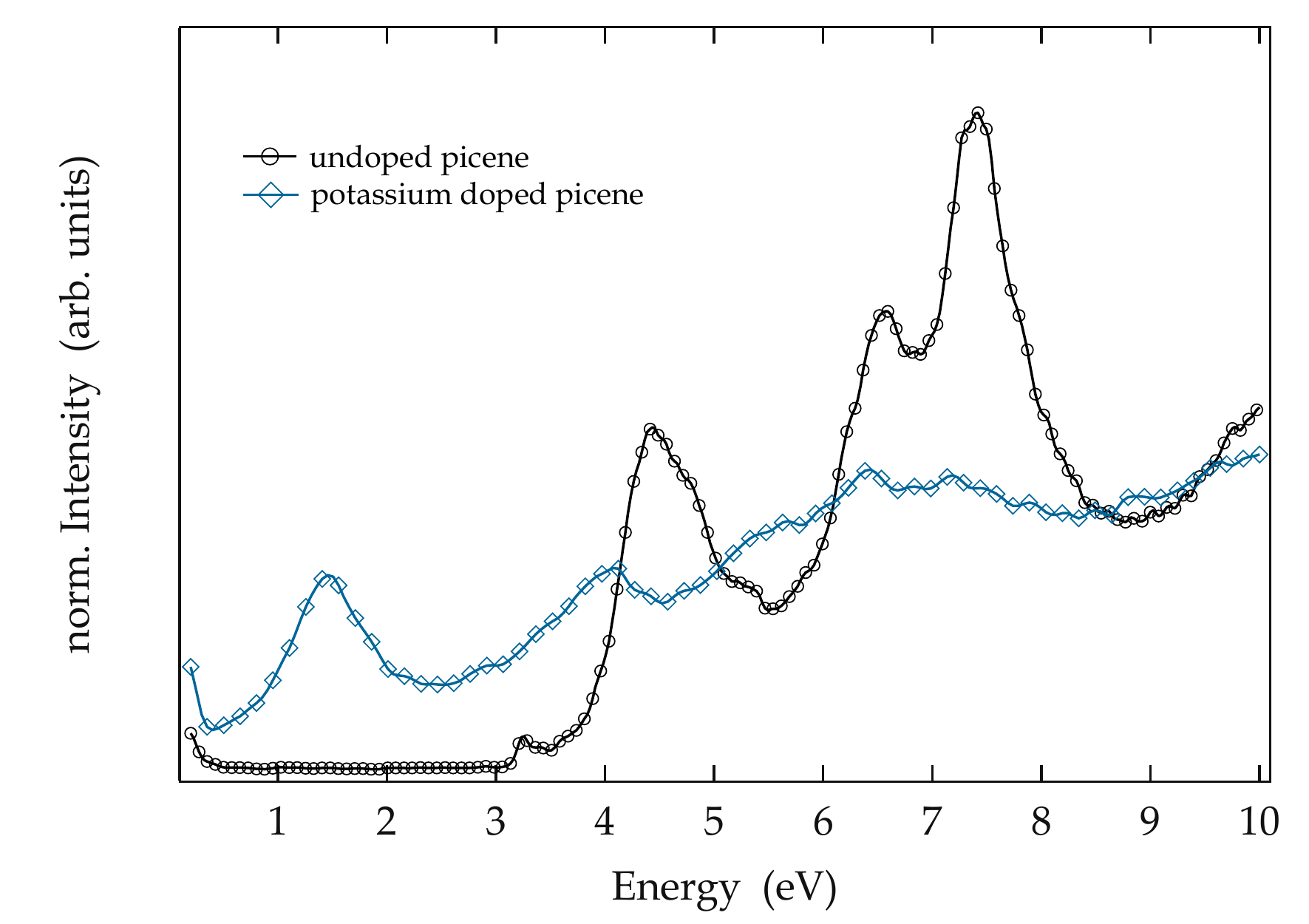}
\includegraphics[width=0.48\linewidth]{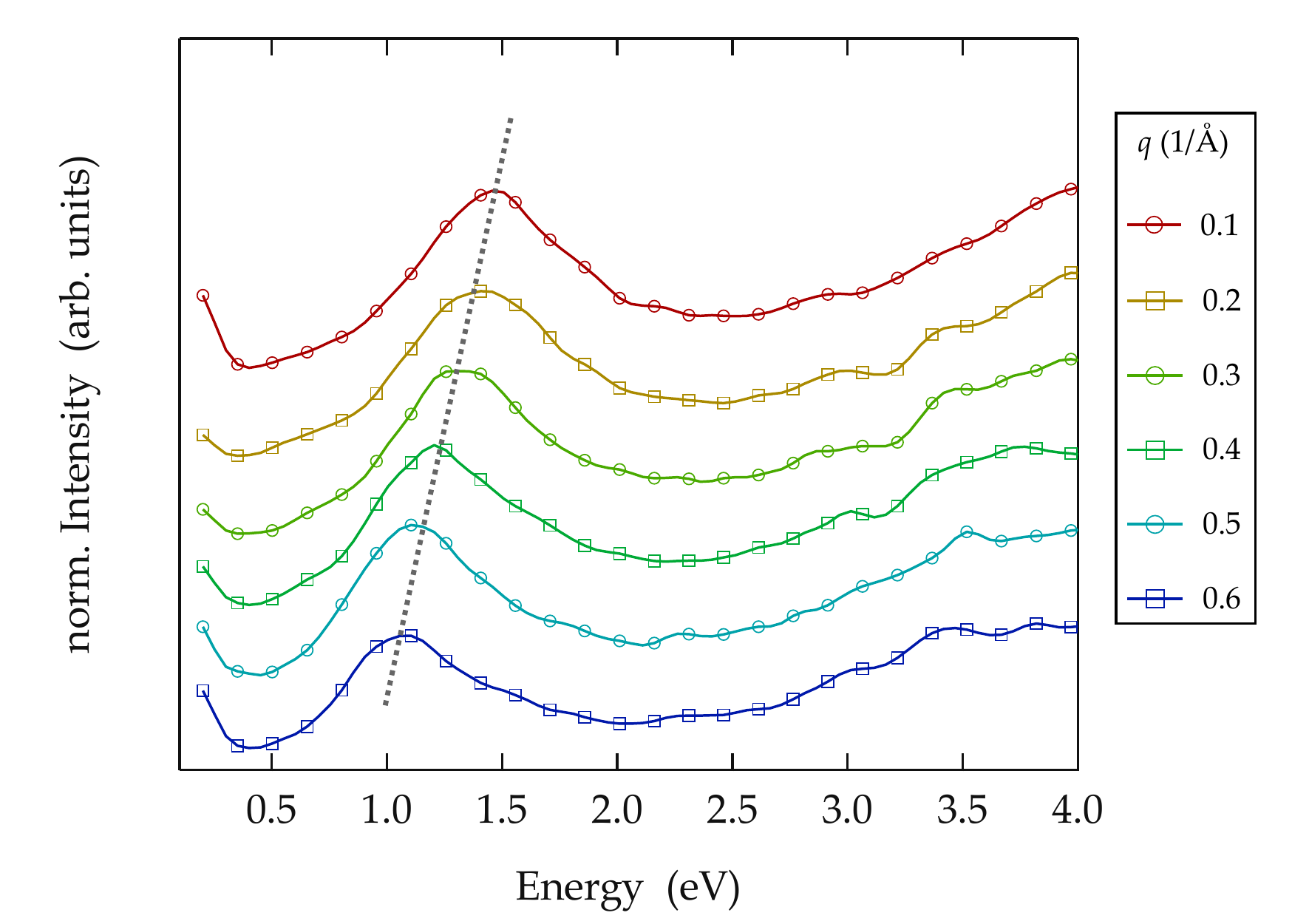}
\caption{Left panel: Loss function of undoped (black circles) and potassium doped (blue diamonds) picene in the range between 0 - 10\,eV. Right panel:  The momentum dependence of the EELS spectra of K$_3$picene ($q$ is increasing from top to bottom spectra).} \label{picene}
\end{figure*}

We have successfully prepared thin films of K$_3$Picene for EELS investigations as described in \cite{Roth2011_c}. Upon potassium addition to picene, various changes in the electronic excitation spectrum are induced. This is outlined in the left panel of Fig.\,\ref{picene}. These data are taken with a small momentum transfer $\boldsymbol{q}$ of 0.1\AA$^{-1}$, which represents the optical limit. For undoped picene, we can clearly identify maxima at about 3.3, 4.6, 6.4, and 7.3\,eV, which are due to excitations between the occupied and unoccupied electronic levels \cite{Roth2010,Roth_2011_b} and which recently have been well reproduced using density functional based calculations \cite{Roth2010}. Note that the spectra in Fig.\,\ref{picene} represent predominantly excitations with a polarization in the $a,b$-crystal plane. Upon doping, the spectral features become broader, and a downshift of the major excitations can be observed. In addition, for the doped film a new structure at about 1.5\,eV is observed in the former gap of picene. Supported by the results of a Kramers Kronig analysis of the measured spectra as describe in \cite{Roth2011}, we can attribute this low energy spectral feature to a collective plasmon excitation of potassium doped picene.

\par

The dispersion of the charge carrier plasmon in K$_3$Picene is presented in the right panel of Fig.\,\ref{picene}. It can be seen that the plasmon dispersion is clearly negative, which deviates from the traditional picture of metals based on the homogeneous electron gas, where the plasmon dispersion should be quadratic and positive (see discussion above). Interestingly, previous investigations of the plasmon dispersion in another doped molecular crystal, K$_3$C$_{60}$, have also revealed an unusual behavior of the plasmon as it is characterized by a vanishing momentum dependence \cite{Gunnarsson1996,Gunnarsson1996_b}.

For molecular solids there exists an important difference to free electron gas metals, which is the strong localization of the conduction electron wave functions to the molecules with little resemblance to a homogeneous distribution. As a consequence, the momentum dependence of the dielectric function and thus the loss function too is different for molecular crystals and those with homogeneous electron gases. In other words, in molecular solids there is strong competition between charge localization and metallicity. This can cause the bare plasmon dispersion for doped molecular solids to be negative \cite{Cudazzo2011}. However, there is even more to be considered for molecular solids. Energetically higher lying inter-band transitions show a quite substantial momentum dependence in their spectral weight, which again is a consequence of the localized wave functions in these materials. As these inter-band transitions represent the dielectric background (screening) of the charge carrier plasmon, they can counteract and even cancel the negative plasmon dispersion \cite{Cudazzo2011,Gunnarsson1996_b}. Finally, since molecular solids have quite inhomogeneous electron systems, also crystal local-field effects have to be considered, which are related to the fact that the polarization due to an external perturbation fluctuates on the atomic scale and the spatial average of the perturbing field is not the same \cite{Cudazzo2011,Gunnarsson1996_b,Kresin1994}. In dependence of the exact crystal packing and molecular structure, in some cases they can even support a negative plasmon dispersion \cite{Cudazzo2011,Gunnarsson1996_b,Kresin1994}.

\subsection{Negative plasmon dispersion in the charge density wave compound \2}
\label{sec:temp-depend-meas}

\2 is a member of the large family of transition metal dichalcogenides, in which many and very interesting physical phenomena occur and can be studied \cite{Wilson1969,Wilson1975,Friend1987}. The general stoichiometry is TX$_2$ with a transition metal T=Ti, Nb, Ta, etc. bound to two chalcogenides X=S, Se, Te. All members show layered structures with only weak van der Waals interlayer interactions and a stacking X-T-X. In addition, the layering allows several geometrical arrangements leading to numerous polymorphs. Depending on the exact composition of the material, various ground states such as semiconductors (e.\,g. MoS$_2$ or WS$_2$), metals (TaSe$_2$, NbSe$_2$), superconductors (e.\,g. NbS$_2$) or charge density waves (e.\,g. TaSe$_2$ or NbSe$_2$) are realized \cite{Wilson1969,Wilson1975}.

\par

\begin{figure}[htbp]
  \centering
  \includegraphics[width=0.48\linewidth]{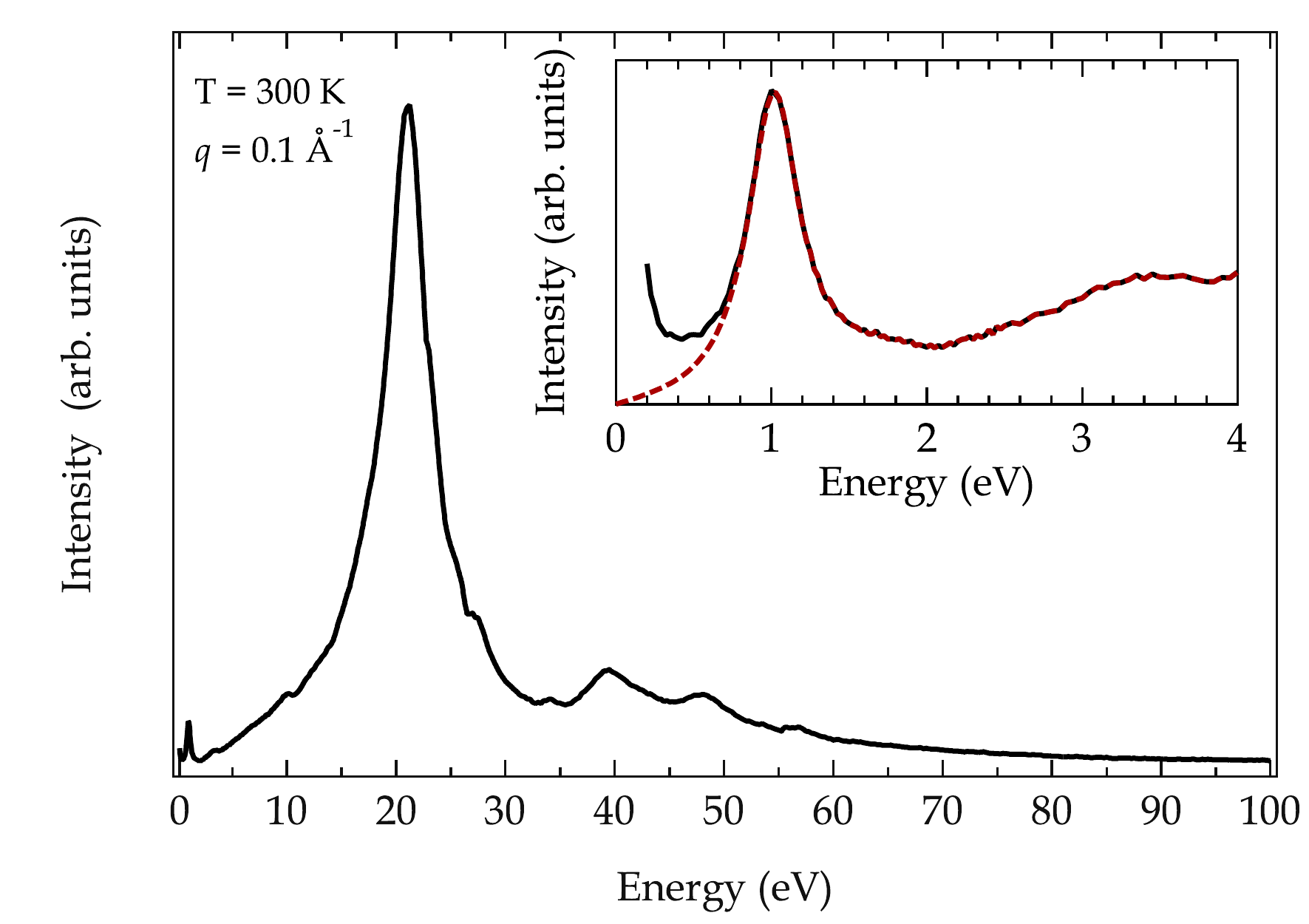}
  \includegraphics[width=0.48\linewidth]{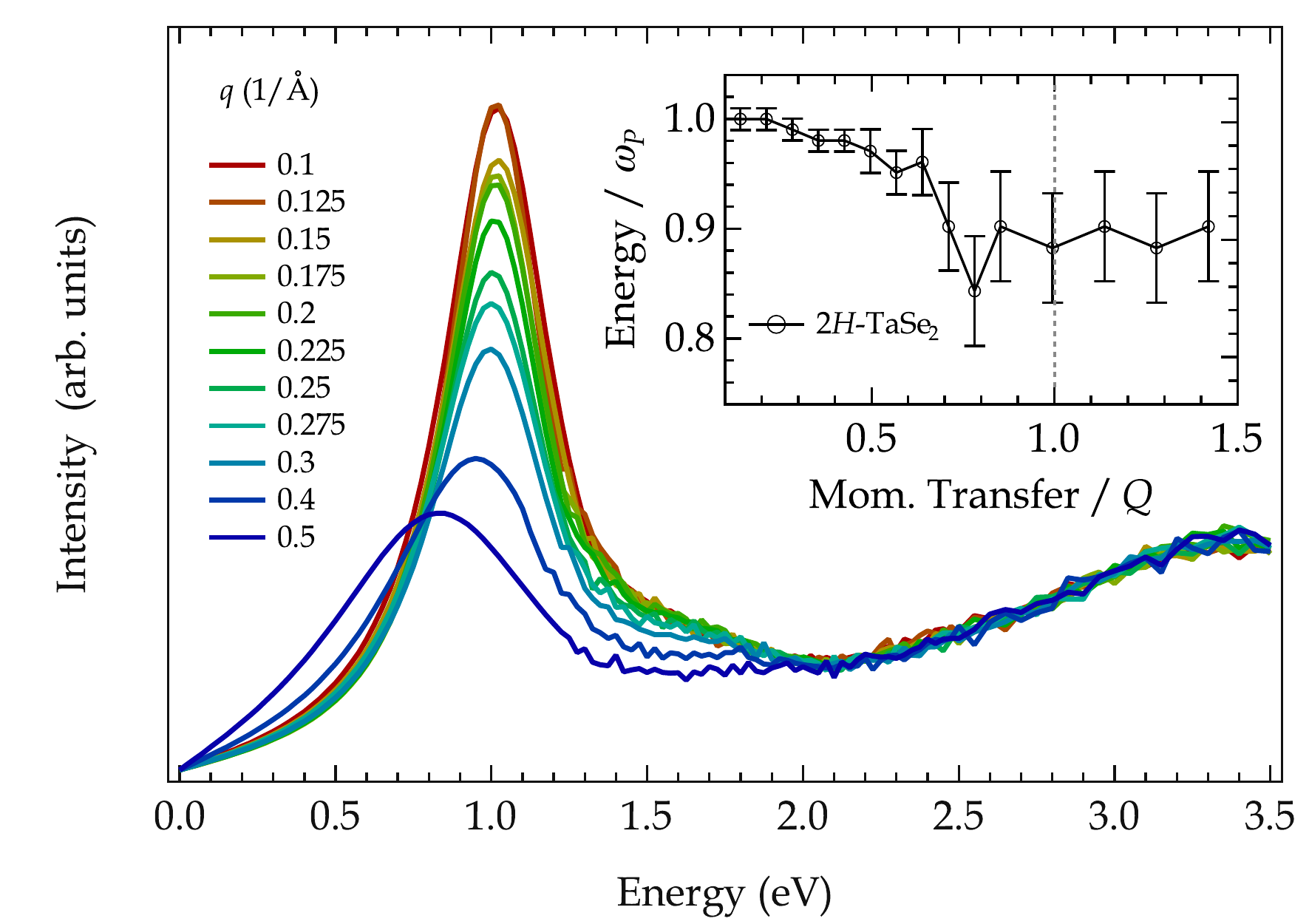}
  \caption{Left panel: Energy-loss spectrum of \2 over a broad energy range, corrected for the elastic line contribution. The inset shows a detailed view of
the low energy regime with the 1\,eV charge-carrier plasmon. Right panel: Loss spectra for \2 at different momentum transfers in the regime of 0.1 to 0.5\AA$^{-1}$. The inset shows the peak position normalised to $\omega_p$ in energy as well as to the CDW ordering vector's absolute value $Q$ in momentum transfer.}
  \label{fig:tase_spectrum}
\end{figure}

The compound discussed in the following, \2, shows a charge density wave (CDW) phase transition by cooling down below 120\,K \cite{Wilson1975,Moncton1975}. The electronic excitations, in particular the behavior of the charge carrier plasmon in \2 have been studied recently using EELS as a function of momentum and temperature. Its energy-loss spectrum in the low energy range is governed by a collective charge carrier plasmon excitation at around 1\,eV (cf.\,Fig.\,\ref{fig:tase_spectrum} left panel). The dominant feature at 21\,eV can be associated to the volume plasmon, while the excitations at even higher energies (around 40\,eV) arise from Ta core-level excitations as well as from multiple scattering of the lower lying excitations.

The behavior of the plasmon as a function of momentum in \2 is very surprising. As can be seen in Fig.\,\ref{fig:tase_spectrum} (right panel) the plasmon dispersion is negative up to a momentum value of about 0.7\,\AA$^{-1}$ where it starts to level off or to increase slightly \cite{vanWezel2011a}. In general, negative plasmon dispersions have been demonstrated in several other materials, among them molecular crystals (see discussion above) but also the heavier alkali metals \cite{Felde1987,Felde1989}.

The presence of a negative plasmon dispersion in \2 with a CDW ground state, and the contrary observation for 2$H$-NbS$_2$ \cite{Manzke1981}, which does not show a phase transition into a low temperature CDW phase, might indicate the importance of charge density wave fluctuations for the momentum dependent behavior of the plasmon. Since the plasmon mode involves a redistribution of electronic charge, it can be strongly coupled to the fluctuations of the charge ordered state, in other words the two density ``waves" compete for the same electrons. In an empirical model put forward recently, the energy required for the excitation of a particular plasmon mode decreases as its wave vector more closely coincides with the preferred wave vector of the charge ordered state and its fluctuations \cite{vanWezel2011a}. The plasmon dispersion is affected whenever the tendency towards electronic order defines a preferred wave vector for the charge density fluctuations. This may thus represent a novel signature of electronic order. It may therefore be expected that the plasmons in a large class of charge ordered materials will develop a signature of the electronic order in their dispersions, in close analogy to that observed in the studied group of transition-metal dichalcogenides.

\par

In contrast, recent calculations of the plasmon dispersion in \2 and related materials based on density functional theory argue in a different direction \cite{Cudazzo2012}. There, the negative plasmon dispersion as observed for \2 is a consequence of the band structure in these compounds, where the conduction electrons occupy bands derived from $d$-levels. It has been concluded that the negative plasmon dispersion results from a peculiar behavior
of the intraband transitions within the $d$-derived conduction bands, that give rise to the plasmon, which are very different from those of a homogeneous electron gas. Since the band structure of many 2$H$ polymorphs of the transition metal dichalcogenides is very similar, a negative plasmon dispersion should be a common behavior of these compounds. This conclusion however contradicts the reported positive plasmon dispersion in 2$H$-NbS$_2$ \cite{Manzke1981}, but it has been argued that this could be due to a non-stoichiometric sample in this case. Therefore, the microscopic origin of the negative plasmon dispersion in \2 still is not understood, and the subject asks for further investigations in order to solve this issue.

\subsection{Unusual plasmon behavior in the spin-ladder cuprate Ca$_{x}$Sr$_{14-x}$Cu$_{24}$O$_{41}$}

Since the discovery of high-temperature superconductivity in Cu-O frameworks \cite{Bednorz1986} a large family of different Cu-O based systems was studied, whereas the dimensionality varied from quasi zero-dimensional systems to two-dimensional networks. The dimension of a system and the associated electronic and magnetic pathway joining neighboring Cu ions, which depends upon the manner in which the CuO$_4$ plaquettes are arranged, plays a key role for the electronic excitations \cite{Knupfer2004_cuprate}.

The family of (La,\,Y,\,Sr,\,Ca)$_{14}$Cu$_{24}$O$_{41}$, the so called spin-ladders, moved into the focus of research after the discovery of superconductivity in the two-leg ladder compound Ca$_{13.6}$Sr$_{0.4}$Cu$_{24}$O$_{41}$ under pressure \cite{Uehara1996}, in particular since this system was the first superconducting copper oxide material with a non-square lattice \cite{Nagata1997}. Interesting is the fact, that superconductivity only occurs when Sr is replaced by Ca. Thereby, the nominal valence of copper remains unchanged but the change in chemical pressure (caused by the different atomic radii of the two elements) within the lattice causes a redistribution of holes from the chain to the ladder subsystem \cite{Kato1996,Nuecker2000,Koitzsch2010}. The compound Ca$_{x}$Sr$_{14-x}$Cu$_{24}$O$_{41}$ is a so-called quasi-one dimensional system and shows additional complexity since it consists of two different types of copper oxide networks---CuO$_2$ (edge sharing) chains and Cu$_2$O$_3$ (two-leg) ladders---which are separated by strings of Sr, Ca and La atoms. These networks are arranged in layers, and the layers are oriented in the crystallographic $ac$-plane while they are stacked in an alternating manner along the perpendicular $b$-axis. Both of these two subsystems, chains and ladders, have orthorhombic symmetry but are structurally incommensurate \cite{McCarron1988,Siegrist1988}.

\begin{figure}[h]
  \centering
  \includegraphics[width=0.45\linewidth]{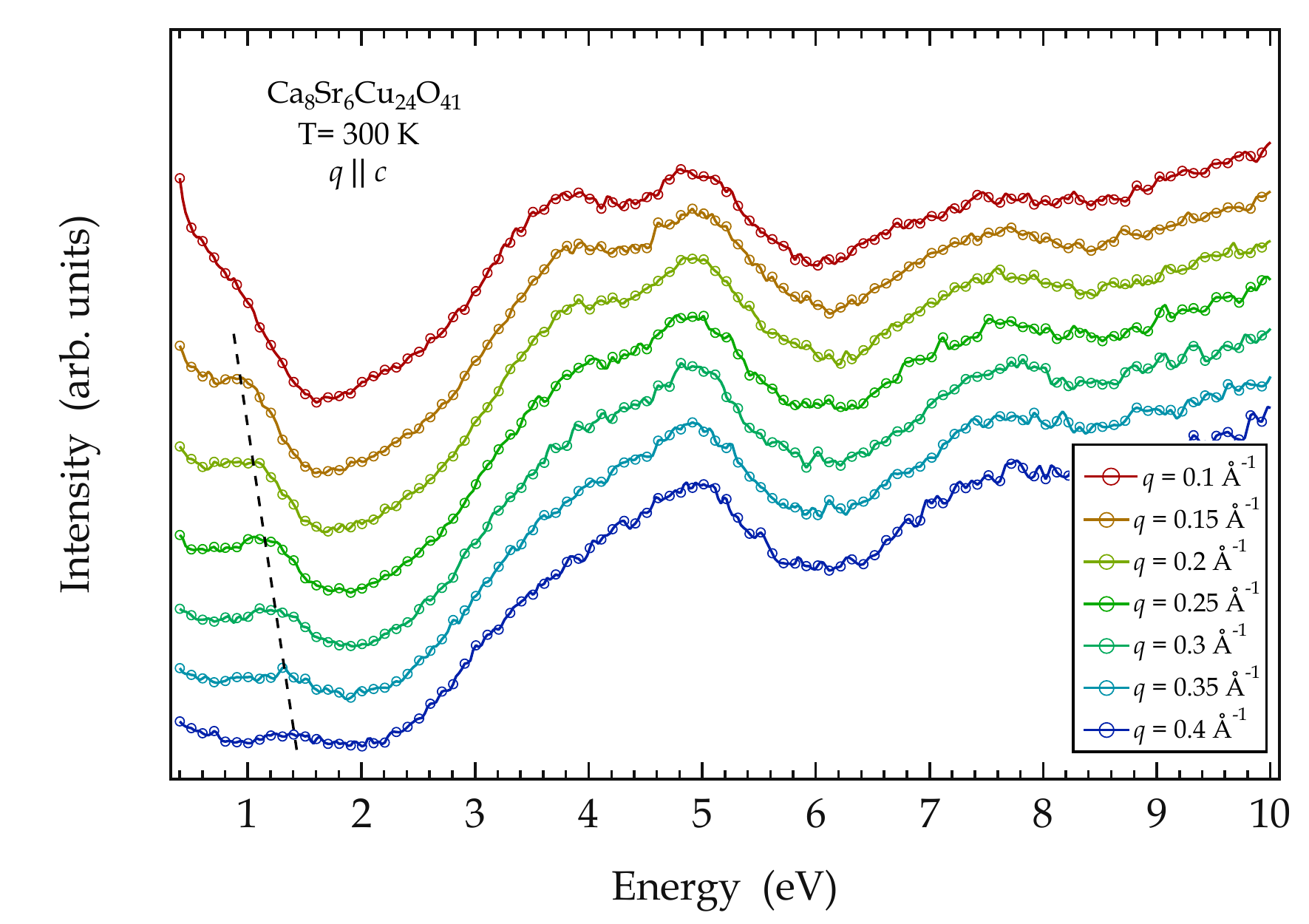}
  \includegraphics[width=0.45\linewidth]{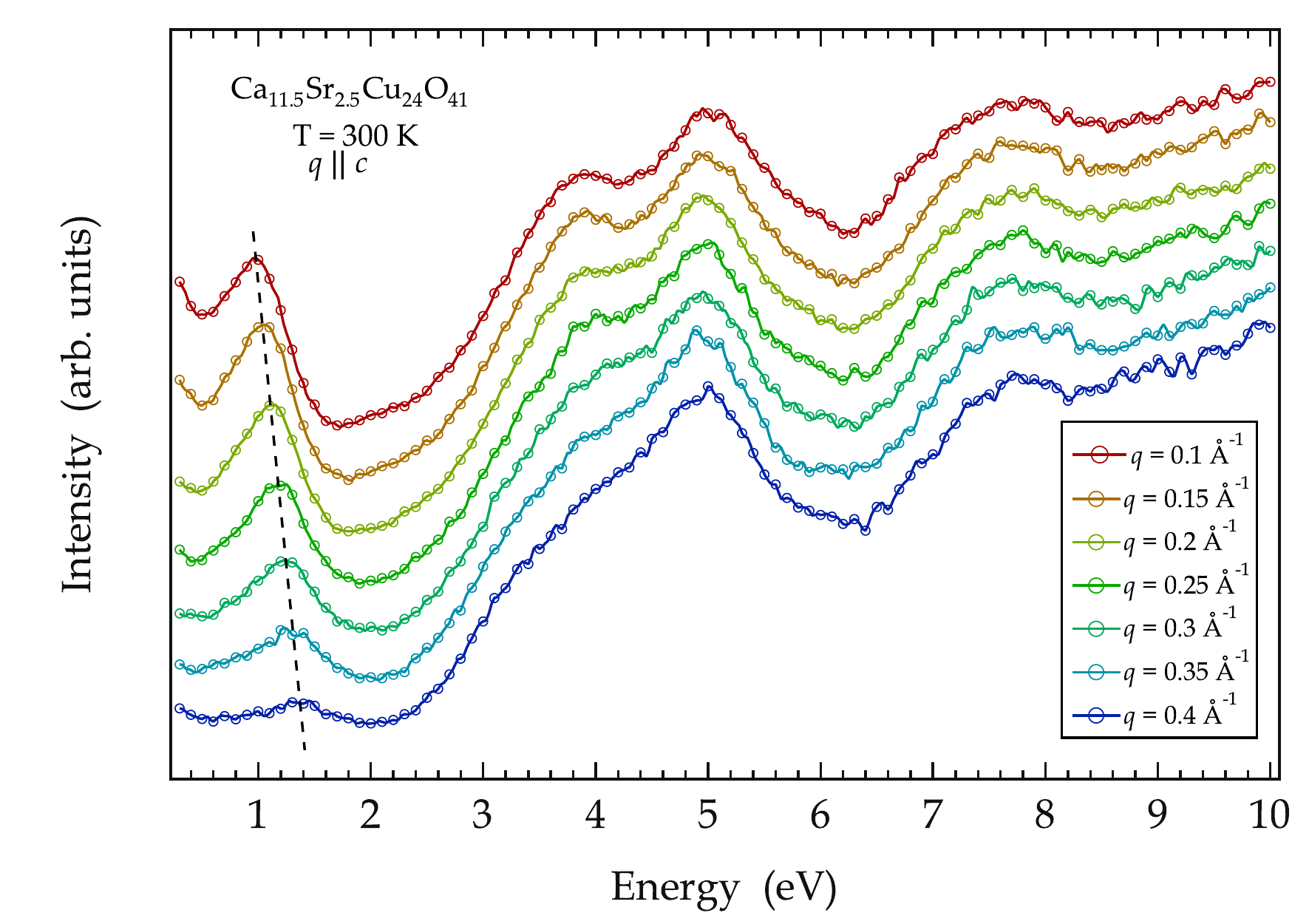}
  \caption{The momentum dependence of the EELS spectra of Ca$_{8}$Sr$_{6}$Cu$_{24}$O$_{41}$ (left panel) and Ca$_{11.5}$Sr$_{2.5}$Cu$_{24}$O$_{41}$ (right panel) for $q$ parallel to the crystallographic $c$\,-\,axis ($q$ is increasing from top to bottom spectra). The upturn towards 0\,eV is due to the quasi-elastic line.}
  \label{fig:Ladder1}
\end{figure}

We have investigated the electronic excitations of Ca$_{x}$Sr$_{14-x}$Cu$_{24}$O$_{41}$ with $x=8$ and $x=11.5$ on thin crystalline films that have been cut from a bulk crystal using an ultra-microtome. Figure\,\ref{fig:Ladder1} displays the evolution of the loss function of Ca$_{8}$Sr$_{6}$Cu$_{24}$O$_{41}$ and Ca$_{11.5}$Sr$_{2.5}$Cu$_{24}$O$_{41}$ with increasing $q$ in an energy range between 0.5\,-\,10\,eV for a momentum transfer parallel to the crystallographic $c$\,-\,axis, i.\,e., parallel to the ladders and chains in the two compounds. The data have been normalized to the high-energy region between 9 and 10\,eV, where they are almost momentum independent. We can clearly identify a well pronounced double peak structure with maxima at
3.5\,-\,4\,eV and at 5\,eV. These spectral features arise from interband transitions in the two compounds, and the similarity in their energy position and momentum behavior suggests a similar origin.

\par

In addition, Fig.\,\ref{fig:Ladder1} reveals a further excitation feature around 1\,eV. In both cases, this additional excitation clearly disperses to higher energies with increasing $q$. Furthermore, the peak width increases with increasing momentum, which indicates damping of this excitation which also is increasing with $q$. According to resistivity data \cite{Motoyama1997}, compounds with higher Ca concentrations show a metallic behavior along  the $c$\,-\,direction, which is also in line with the appearance of a plasma edge close around 1\,eV in the corresponding reflectivity spectra \cite{Osafune1997,Ruzicka1998}. Consequently, we ascribe the peak around 1\,eV to the charge carrier plasmon of Ca$_{8}$Sr$_{9}$Cu$_{24}$O$_{41}$ and Ca$_{11.5}$Sr$_{2.5}$Cu$_{24}$O$_{41}$, respectively, which is polarized along the ladder direction. We note that the plasmon energy is analogous to what has been observed for other similarly doped, two-dimensional cuprate systems \cite{Wang1990,Nuecker1991,Knupfer1994}.

\begin{figure*}
  \centering
  \includegraphics[width=0.45\linewidth]{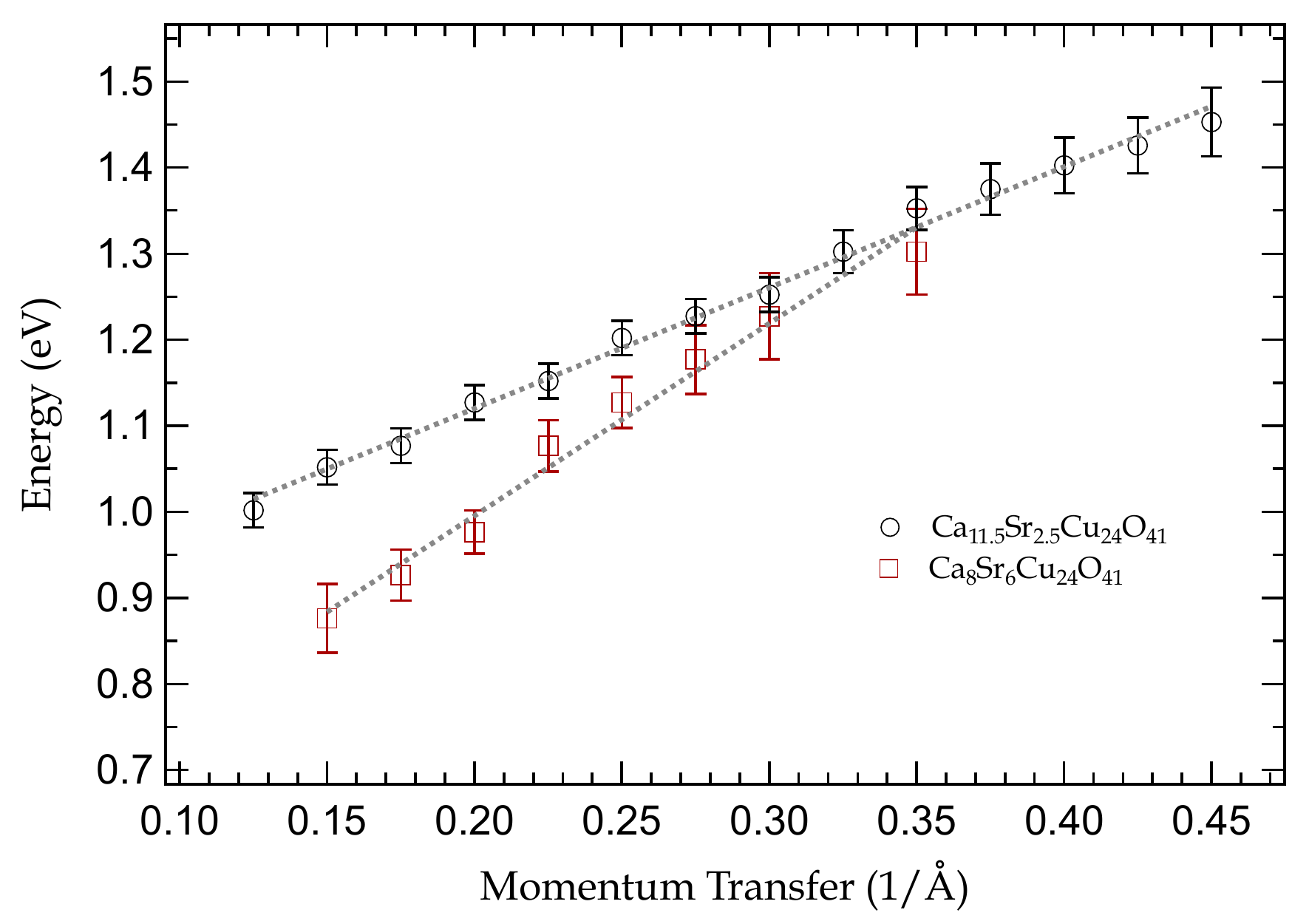}
  \caption{Comparison of the plasmon dispersion in Ca$_{8}$Sr$_{9}$Cu$_{24}$O$_{41}$ (black squares) and Ca$_{11.5}$Sr$_{2.5}$Cu$_{24}$O$_{41}$ (red circles) along the $c$\,-\,direction.}
  \label{fig:Ladder2}
\end{figure*}

In order to further quantify the behavior of the charge carrier plasmon we present in Fig.\,\ref{fig:Ladder2} the evolution of the peak position in the range 0.125\,\AA$^{-1}$ to 0.45\,\AA$^{-1}$ for $q \parallel$ $c$. Due to the strong damping of the plasmon and the low cross section for higher momentum transfers, data for a momentum transfer above $q$\,=\,0.35\,\AA$^{-1}$ ($x=8$) and above $q$\,=\,0.45\,\AA$^{-1}$ ($x=11.5$) are not included.

\par

Fig.\,\ref{fig:Ladder2} provides several interesting aspects on the plasmons in Ca$_{x}$Sr$_{14-x}$Cu$_{24}$O$_{41}$ compounds. Upon increasing Ca content, the plasmon energy shift to higher values. This is expected since higher Ca content results in a higher charge carrier concentration in the CuO$_2$ ladders within these materials. From the measured plasmon energies our data allow a rough estimate of the increase in charge carrier density ($\omega_p^2 \propto n^2$) of about a factor of 1.4 to 1.5 going from Ca$_{8}$Sr$_{6}$Cu$_{24}$O$_{41}$ to Ca$_{11.5}$Sr$_{2.5}$Cu$_{24}$O$_{41}$. We note however, that this estimate requires that both the background dielectric constant and the effective mass of the charge carriers is not or only weakly doping dependent, an assumption that is probably invalid for Ca$_{x}$Sr$_{14-x}$Cu$_{24}$O$_{41}$.

\par

In Fig.\,\ref{fig:Ladder2} it can also be seen that the plasmon dispersion for both compounds is positive, in the case of Ca$_{11.5}$Sr$_{2.5}$Cu$_{24}$O$_{41}$ the measured dispersion range of about 500\,meV corresponds to what has been reported for planar, optimally doped cuprates such as Bi$_2$Sr$_2$CaCu$_2$O$_{8-\delta}$ \cite{Nuecker1991}. The latter seems reasonable, taking into account that the doping level of the ladders in Ca$_{11.5}$Sr$_{2.5}$Cu$_{24}$O$_{41}$ is about 0.15 - 0.2 holes per Cu unit, as reported from recent angular resolved photoemission experiments \cite{Koitzsch2010}, i.\,e., of similar size as in the optimally doped Bi$_2$Sr$_2$CaCu$_2$O$_{8-\delta}$. Interestingly, the slope of the plasmon dispersion is significantly doping dependent. It decreases when going from Ca$_{8}$Sr$_{6}$Cu$_{24}$O$_{41}$ to Ca$_{11.5}$Sr$_{2.5}$Cu$_{24}$O$_{41}$, while the plasmon energy (at low momenta) increases. This is in contradiction to what would be expected for a homogeneous electron gas. There the plasmon energy $\omega_p(q=0)$ increases proportional to $n^{1/2}$ ($n$ is the charge carrier density), but also the slope of the plasmon dispersion would increase proportional to $n^{1/6}$ \cite{Raether1979}. This clearly signals that plasmons in correlated materials cannot be described using the simple electron gas approach. Rather, the changes in the electronic band structure and most likely also electronic correlation effects play a considerable role. This conclusion is further supported by the observation of a quasi-linear plasmon dispersion in Fig.\,\ref{fig:Ladder2}. In this context we emphasize that for the planar, two-dimensional Bi$_2$Sr$_2$CaCu$_2$O$_{8-\delta}$ a quadratic plasmon dispersion has been reported, which points to an important role of the dimensionality of the structures.

\par

Deviations from the expectation of a quadratic plasmon dispersion have already been reported in the past and also in the chapters above. In regard of
quasi one-dimensional metallic systems, there are studies for a few compounds. Within RPA it has been predicted \cite{Williams1974} that the plasmon dispersion in one dimensional metals can be substantially modified by local field effects, i.\,e., the inhomogeneous character of the electron gas. This modification can even cause a negative plasmon dispersion in case of a tight binding description of the electronic bands \cite{Williams1974}.
Experimental studies of the plasmon dispersion in quasi one-dimensional (TaSe$_4$)$_2$I \cite{Sing1998} and K$_{0.3}$MoO$_3$, \cite{Sing1999} found a quasi-linear dispersion, which in these cases has been explained to predominantly be an effect of the band structure in these materials.

\par

Moreover, going to lower doping levels of about 0.1 holes per Cu atom in two-dimensional cuprate structures, the plasmon dispersion is  drastically reduced compared to optimal doping (i.\,e. about 0.15 holes per copper atom). The band width of the plasmon in
Ca$_{1.9}$Na$_{0.1}$CuO$_2$Cl$_2$ is only half of that observed for the doping level of 0.15 holes per copper unit \cite{Schuster2010}.
And finally, for Ca$_{x}$Sr$_{14-x}$Cu$_{24}$O$_{41}$ compounds the formation of a hole crystal \cite{Abbamonte2004,Carr2002,Friedel2002,Hess2004} (i.\,e., a charge density wave) has been reported. These findings suggest that also in the cuprate ladders there might be a phase quite similar to the pseudo gap phase in the planar cuprates, with complex electronic degrees of freedom and interactions. To this end, a final rationalization of the behavior of plasmons in the quasi one-dimensional correlated Ca$_{x}$Sr$_{14-x}$Cu$_{24}$O$_{41}$ compounds still is not achieved.

\section{Summary}

To summarize, we have demonstrated that energy and momentum of plasmons can depend on other degrees of freedom in the material under investigation. In turn, a detailed study of the plasmon behavior allows insight into the complex behavior of materials, and thus contributes to the development of our fundamental knowledge of the electronic structure of condensed matter. The very important improvements of the EELS technique in the last years, especially the increased energy resolution, helps to answer current questions in solid state physics and leads to new and sometime unexpected insights into the electronic structure of actual and novel materials.

\section{Acknowledgments}
We thank M. Naumann, R. Sch\"onfelder, R. H\"ubel and S. Leger for technical assistance. Part of this work has been supported by the Deutsche Forschungsgemeinschaft under KN393/13, KN393/14 and GRK1621.


\begin{thebibliography}{96}%
\makeatletter
\providecommand \@ifxundefined [1]{%
 \@ifx{#1\undefined}
}%
\providecommand \@ifnum [1]{%
 \ifnum #1\expandafter \@firstoftwo
 \else \expandafter \@secondoftwo
 \fi
}%
\providecommand \@ifx [1]{%
 \ifx #1\expandafter \@firstoftwo
 \else \expandafter \@secondoftwo
 \fi
}%
\providecommand \natexlab [1]{#1}%
\providecommand \enquote  [1]{``#1''}%
\providecommand \bibnamefont  [1]{#1}%
\providecommand \bibfnamefont [1]{#1}%
\providecommand \citenamefont [1]{#1}%
\providecommand \href@noop [0]{\@secondoftwo}%
\providecommand \href [0]{\begingroup \@sanitize@url \@href}%
\providecommand \@href[1]{\@@startlink{#1}\@@href}%
\providecommand \@@href[1]{\endgroup#1\@@endlink}%
\providecommand \@sanitize@url [0]{\catcode `\\12\catcode `\$12\catcode
  `\&12\catcode `\#12\catcode `\^12\catcode `\_12\catcode `\%12\relax}%
\providecommand \@@startlink[1]{}%
\providecommand \@@endlink[0]{}%
\providecommand \url  [0]{\begingroup\@sanitize@url \@url }%
\providecommand \@url [1]{\endgroup\@href {#1}{\urlprefix }}%
\providecommand \urlprefix  [0]{URL }%
\providecommand \Eprint [0]{\href }%
\providecommand \doibase [0]{http://dx.doi.org/}%
\providecommand \selectlanguage [0]{\@gobble}%
\providecommand \bibinfo  [0]{\@secondoftwo}%
\providecommand \bibfield  [0]{\@secondoftwo}%
\providecommand \translation [1]{[#1]}%
\providecommand \BibitemOpen [0]{}%
\providecommand \bibitemStop [0]{}%
\providecommand \bibitemNoStop [0]{.\EOS\space}%
\providecommand \EOS [0]{\spacefactor3000\relax}%
\providecommand \BibitemShut  [1]{\csname bibitem#1\endcsname}%
\let\auto@bib@innerbib\@empty
\bibitem [{\citenamefont {Raether}(1980)}]{Raether1979}%
  \BibitemOpen
  \bibfield  {author} {\bibinfo {author} {\bibfnamefont {H.}~\bibnamefont
  {Raether}},\ }\href@noop {} {\emph {\bibinfo {title} {Excitation of plasmons
  and interband transitions by electrons}}}\ (\bibinfo  {publisher} {Springer
  Verlag, Berlin},\ \bibinfo {year} {1980})\BibitemShut {NoStop}%
\bibitem [{\citenamefont {Schnatterly}(1979)}]{Schnatterly1979}%
  \BibitemOpen
  \bibfield  {author} {\bibinfo {author} {\bibfnamefont {S.}~\bibnamefont
  {Schnatterly}}\ }(\bibinfo  {publisher} {Academic Press},\ \bibinfo {year}
  {1979})\ pp.\ \bibinfo {pages} {275 -- 358}\BibitemShut {NoStop}%
\bibitem [{\citenamefont {Lucas}\ and\ \citenamefont
  {\v{S}unji\'{c}}(1972)}]{Lucas1972}%
  \BibitemOpen
  \bibfield  {author} {\bibinfo {author} {\bibfnamefont {A.}~\bibnamefont
  {Lucas}}\ and\ \bibinfo {author} {\bibfnamefont {M.}~\bibnamefont
  {\v{S}unji\'{c}}},\ }\href {\doibase
  http://dx.doi.org/10.1016/0079-6816(72)90002-0} {\bibfield  {journal}
  {\bibinfo  {journal} {Prog. Surf. Sci.}\ }\textbf {\bibinfo {volume} {2, Part
  2}},\ \bibinfo {pages} {75 } (\bibinfo {year} {1972})}\BibitemShut {NoStop}%
\bibitem [{\citenamefont {Sch\"ulke}(2007)}]{Schuelke2007}%
  \BibitemOpen
  \bibfield  {author} {\bibinfo {author} {\bibfnamefont {W.}~\bibnamefont
  {Sch\"ulke}},\ }\href@noop {} {\emph {\bibinfo {title} {Electron Dynamics by
  Inelastic X-Ray Scattering}}}\ (\bibinfo  {publisher} {Oxford University
  Press, USA},\ \bibinfo {year} {2007})\BibitemShut {NoStop}%
\bibitem [{\citenamefont {Kotani}\ and\ \citenamefont
  {Shin}(2001)}]{Kotani2001}%
  \BibitemOpen
  \bibfield  {author} {\bibinfo {author} {\bibfnamefont {A.}~\bibnamefont
  {Kotani}}\ and\ \bibinfo {author} {\bibfnamefont {S.}~\bibnamefont {Shin}},\
  }\href {\doibase 10.1103/RevModPhys.73.203} {\bibfield  {journal} {\bibinfo
  {journal} {Rev. Mod. Phys.}\ }\textbf {\bibinfo {volume} {73}},\ \bibinfo
  {pages} {203} (\bibinfo {year} {2001})}\BibitemShut {NoStop}%
\bibitem [{\citenamefont {Ament}\ \emph {et~al.}(2011)\citenamefont {Ament},
  \citenamefont {van Veenendaal}, \citenamefont {Devereaux}, \citenamefont
  {Hill},\ and\ \citenamefont {van~den Brink}}]{Ament2011}%
  \BibitemOpen
  \bibfield  {author} {\bibinfo {author} {\bibfnamefont {L.~J.~P.}\
  \bibnamefont {Ament}}, \bibinfo {author} {\bibfnamefont {M.}~\bibnamefont
  {van Veenendaal}}, \bibinfo {author} {\bibfnamefont {T.~P.}\ \bibnamefont
  {Devereaux}}, \bibinfo {author} {\bibfnamefont {J.~P.}\ \bibnamefont {Hill}},
  \ and\ \bibinfo {author} {\bibfnamefont {J.}~\bibnamefont {van~den Brink}},\
  }\href {\doibase 10.1103/RevModPhys.83.705} {\bibfield  {journal} {\bibinfo
  {journal} {Rev. Mod. Phys.}\ }\textbf {\bibinfo {volume} {83}},\ \bibinfo
  {pages} {705} (\bibinfo {year} {2011})}\BibitemShut {NoStop}%
\bibitem [{\citenamefont {Egerton}(1996)}]{Egertonbuch}%
  \BibitemOpen
  \bibfield  {author} {\bibinfo {author} {\bibfnamefont {R.}~\bibnamefont
  {Egerton}},\ }\href@noop {} {\emph {\bibinfo {title} {Electron Energy-Loss
  Spectroscopy in the Electron Microscope}}},\ \bibinfo {edition} {2nd}\ ed.\
  (\bibinfo  {publisher} {Plenum Press, New York},\ \bibinfo {year}
  {1996})\BibitemShut {NoStop}%
\bibitem [{\citenamefont {Egerton}(2009)}]{Egerton2009}%
  \BibitemOpen
  \bibfield  {author} {\bibinfo {author} {\bibfnamefont {R.~F.}\ \bibnamefont
  {Egerton}},\ }\href {http://stacks.iop.org/0034-4885/72/i=1/a=016502}
  {\bibfield  {journal} {\bibinfo  {journal} {Rep. Prog. Phys.}\ }\textbf
  {\bibinfo {volume} {72}},\ \bibinfo {pages} {016502} (\bibinfo {year}
  {2009})}\BibitemShut {NoStop}%
\bibitem [{\citenamefont {Garcia~de Abajo}(2010)}]{Abajo2010}%
  \BibitemOpen
  \bibfield  {author} {\bibinfo {author} {\bibfnamefont {F.~J.}\ \bibnamefont
  {Garcia~de Abajo}},\ }\href {\doibase 10.1103/RevModPhys.82.209} {\bibfield
  {journal} {\bibinfo  {journal} {Rev. Mod. Phys.}\ }\textbf {\bibinfo {volume}
  {82}},\ \bibinfo {pages} {209} (\bibinfo {year} {2010})}\BibitemShut
  {NoStop}%
\bibitem [{\citenamefont {Schattschneider}(1986)}]{Schattschneider1986}%
  \BibitemOpen
  \bibfield  {author} {\bibinfo {author} {\bibfnamefont {P.}~\bibnamefont
  {Schattschneider}},\ }\href@noop {} {\emph {\bibinfo {title} {Fundamentals of
  inelastic electron scattering}}}\ (\bibinfo  {publisher} {Springer},\
  \bibinfo {year} {1986})\BibitemShut {NoStop}%
\bibitem [{\citenamefont {Ibach}\ and\ \citenamefont
  {Mills}(1982)}]{Ibach1982}%
  \BibitemOpen
  \bibfield  {author} {\bibinfo {author} {\bibfnamefont {H.}~\bibnamefont
  {Ibach}}\ and\ \bibinfo {author} {\bibfnamefont {D.~L.}\ \bibnamefont
  {Mills}},\ }\href@noop {} {\emph {\bibinfo {title} {Electron Energy Loss
  Spectroscopy and Surface Vibrations}}}\ (\bibinfo  {publisher} {Academic
  Press, New York},\ \bibinfo {year} {1982})\BibitemShut {NoStop}%
\bibitem [{\citenamefont {Ibach}(1993)}]{Ibach1993}%
  \BibitemOpen
  \bibfield  {author} {\bibinfo {author} {\bibfnamefont {H.}~\bibnamefont
  {Ibach}},\ }\href {\doibase http://dx.doi.org/10.1016/0368-2048(93)80155-F}
  {\bibfield  {journal} {\bibinfo  {journal} {J. Electron. Spectrosc. Relat.
  Phenom.}\ }\textbf {\bibinfo {volume} {64-65}},\ \bibinfo {pages} {819 }
  (\bibinfo {year} {1993})}\BibitemShut {NoStop}%
\bibitem [{\citenamefont {Richardson}(1997)}]{Richardson1997}%
  \BibitemOpen
  \bibfield  {author} {\bibinfo {author} {\bibfnamefont {N.~V.}\ \bibnamefont
  {Richardson}},\ }\href {\doibase
  http://dx.doi.org/10.1016/S1359-0286(97)80039-3} {\bibfield  {journal}
  {\bibinfo  {journal} {Curr. Opin. Solid State Mater. Sci.}\ }\textbf
  {\bibinfo {volume} {2}},\ \bibinfo {pages} {517 } (\bibinfo {year}
  {1997})}\BibitemShut {NoStop}%
\bibitem [{\citenamefont {Fink}(1989)}]{Fink1989}%
  \BibitemOpen
  \bibfield  {author} {\bibinfo {author} {\bibfnamefont {J.}~\bibnamefont
  {Fink}},\ }\href
  {http://www.sciencedirect.com/science/article/pii/S0065253908609476}
  {\bibfield  {journal} {\bibinfo  {journal} {Adv. Electr. Electr. Phys.}\
  }\textbf {\bibinfo {volume} {75}},\ \bibinfo {pages} {121} (\bibinfo {year}
  {1989})}\BibitemShut {NoStop}%
\bibitem [{\citenamefont {Pichler}\ \emph {et~al.}(1998)\citenamefont
  {Pichler}, \citenamefont {Knupfer}, \citenamefont {Golden}, \citenamefont
  {Fink}, \citenamefont {Rinzler},\ and\ \citenamefont
  {Smalley}}]{Pichler1998}%
  \BibitemOpen
  \bibfield  {author} {\bibinfo {author} {\bibfnamefont {T.}~\bibnamefont
  {Pichler}}, \bibinfo {author} {\bibfnamefont {M.}~\bibnamefont {Knupfer}},
  \bibinfo {author} {\bibfnamefont {M.~S.}\ \bibnamefont {Golden}}, \bibinfo
  {author} {\bibfnamefont {J.}~\bibnamefont {Fink}}, \bibinfo {author}
  {\bibfnamefont {A.}~\bibnamefont {Rinzler}}, \ and\ \bibinfo {author}
  {\bibfnamefont {R.~E.}\ \bibnamefont {Smalley}},\ }\href {\doibase
  10.1103/PhysRevLett.80.4729} {\bibfield  {journal} {\bibinfo  {journal}
  {Phys. Rev. Lett.}\ }\textbf {\bibinfo {volume} {80}},\ \bibinfo {pages}
  {4729} (\bibinfo {year} {1998})}\BibitemShut {NoStop}%
\bibitem [{\citenamefont {Neudert}\ \emph {et~al.}(1998)\citenamefont
  {Neudert}, \citenamefont {Knupfer}, \citenamefont {Golden}, \citenamefont
  {Fink}, \citenamefont {Stephan}, \citenamefont {Penc}, \citenamefont
  {Motoyama}, \citenamefont {Eisaki},\ and\ \citenamefont
  {Uchida}}]{Neudert1998}%
  \BibitemOpen
  \bibfield  {author} {\bibinfo {author} {\bibfnamefont {R.}~\bibnamefont
  {Neudert}}, \bibinfo {author} {\bibfnamefont {M.}~\bibnamefont {Knupfer}},
  \bibinfo {author} {\bibfnamefont {M.~S.}\ \bibnamefont {Golden}}, \bibinfo
  {author} {\bibfnamefont {J.}~\bibnamefont {Fink}}, \bibinfo {author}
  {\bibfnamefont {W.}~\bibnamefont {Stephan}}, \bibinfo {author} {\bibfnamefont
  {K.}~\bibnamefont {Penc}}, \bibinfo {author} {\bibfnamefont {N.}~\bibnamefont
  {Motoyama}}, \bibinfo {author} {\bibfnamefont {H.}~\bibnamefont {Eisaki}}, \
  and\ \bibinfo {author} {\bibfnamefont {S.}~\bibnamefont {Uchida}},\ }\href
  {\doibase 10.1103/PhysRevLett.81.657} {\bibfield  {journal} {\bibinfo
  {journal} {Phys. Rev. Lett.}\ }\textbf {\bibinfo {volume} {81}},\ \bibinfo
  {pages} {657} (\bibinfo {year} {1998})}\BibitemShut {NoStop}%
\bibitem [{\citenamefont {Knupfer}\ \emph {et~al.}(1999)\citenamefont
  {Knupfer}, \citenamefont {Pichler}, \citenamefont {Golden}, \citenamefont
  {Fink}, \citenamefont {Murgia}, \citenamefont {Michel}, \citenamefont
  {Zamboni},\ and\ \citenamefont {Taliani}}]{Knupfer1999_2}%
  \BibitemOpen
  \bibfield  {author} {\bibinfo {author} {\bibfnamefont {M.}~\bibnamefont
  {Knupfer}}, \bibinfo {author} {\bibfnamefont {T.}~\bibnamefont {Pichler}},
  \bibinfo {author} {\bibfnamefont {M.~S.}\ \bibnamefont {Golden}}, \bibinfo
  {author} {\bibfnamefont {J.}~\bibnamefont {Fink}}, \bibinfo {author}
  {\bibfnamefont {M.}~\bibnamefont {Murgia}}, \bibinfo {author} {\bibfnamefont
  {R.~H.}\ \bibnamefont {Michel}}, \bibinfo {author} {\bibfnamefont
  {R.}~\bibnamefont {Zamboni}}, \ and\ \bibinfo {author} {\bibfnamefont
  {C.}~\bibnamefont {Taliani}},\ }\href {\doibase 10.1103/PhysRevLett.83.1443}
  {\bibfield  {journal} {\bibinfo  {journal} {Phys. Rev. Lett.}\ }\textbf
  {\bibinfo {volume} {83}},\ \bibinfo {pages} {1443} (\bibinfo {year}
  {1999})}\BibitemShut {NoStop}%
\bibitem [{\citenamefont {Knupfer}\ \emph {et~al.}(2000)\citenamefont
  {Knupfer}, \citenamefont {Fink}, \citenamefont {Zojer}, \citenamefont
  {Leising},\ and\ \citenamefont {Fichou}}]{Knupfer2005}%
  \BibitemOpen
  \bibfield  {author} {\bibinfo {author} {\bibfnamefont {M.}~\bibnamefont
  {Knupfer}}, \bibinfo {author} {\bibfnamefont {J.}~\bibnamefont {Fink}},
  \bibinfo {author} {\bibfnamefont {E.}~\bibnamefont {Zojer}}, \bibinfo
  {author} {\bibfnamefont {G.}~\bibnamefont {Leising}}, \ and\ \bibinfo
  {author} {\bibfnamefont {D.}~\bibnamefont {Fichou}},\ }\href {\doibase
  http://dx.doi.org/10.1016/S0009-2614(00)00033-6} {\bibfield  {journal}
  {\bibinfo  {journal} {Chem. Phys. Lett.}\ }\textbf {\bibinfo {volume}
  {318}},\ \bibinfo {pages} {585 } (\bibinfo {year} {2000})}\BibitemShut
  {NoStop}%
\bibitem [{\citenamefont {Knupfer}\ \emph {et~al.}(2002)\citenamefont
  {Knupfer}, \citenamefont {Schwieger}, \citenamefont {Fink}, \citenamefont
  {Leo},\ and\ \citenamefont {Hoffmann}}]{Knupfer2002}%
  \BibitemOpen
  \bibfield  {author} {\bibinfo {author} {\bibfnamefont {M.}~\bibnamefont
  {Knupfer}}, \bibinfo {author} {\bibfnamefont {T.}~\bibnamefont {Schwieger}},
  \bibinfo {author} {\bibfnamefont {J.}~\bibnamefont {Fink}}, \bibinfo {author}
  {\bibfnamefont {K.}~\bibnamefont {Leo}}, \ and\ \bibinfo {author}
  {\bibfnamefont {M.}~\bibnamefont {Hoffmann}},\ }\href {\doibase
  10.1103/PhysRevB.66.035208} {\bibfield  {journal} {\bibinfo  {journal} {Phys.
  Rev. B}\ }\textbf {\bibinfo {volume} {66}},\ \bibinfo {pages} {035208}
  (\bibinfo {year} {2002})}\BibitemShut {NoStop}%
\bibitem [{\citenamefont {Schuster}\ \emph {et~al.}(2007)\citenamefont
  {Schuster}, \citenamefont {Knupfer},\ and\ \citenamefont
  {Berger}}]{Schuster2007_penta}%
  \BibitemOpen
  \bibfield  {author} {\bibinfo {author} {\bibfnamefont {R.}~\bibnamefont
  {Schuster}}, \bibinfo {author} {\bibfnamefont {M.}~\bibnamefont {Knupfer}}, \
  and\ \bibinfo {author} {\bibfnamefont {H.}~\bibnamefont {Berger}},\ }\href
  {\doibase 10.1103/PhysRevLett.98.037402} {\bibfield  {journal} {\bibinfo
  {journal} {Phys. Rev. Lett.}\ }\textbf {\bibinfo {volume} {98}},\ \bibinfo
  {pages} {037402} (\bibinfo {year} {2007})}\BibitemShut {NoStop}%
\bibitem [{\citenamefont {Kramberger}\ \emph {et~al.}(2008)\citenamefont
  {Kramberger}, \citenamefont {Hambach}, \citenamefont {Giorgetti},
  \citenamefont {R\"ummeli}, \citenamefont {Knupfer}, \citenamefont {Fink},
  \citenamefont {B\"uchner}, \citenamefont {Reining}, \citenamefont
  {Einarsson}, \citenamefont {Maruyama}, \citenamefont {Sottile}, \citenamefont
  {Hannewald}, \citenamefont {Olevano}, \citenamefont {Marinopoulos},\ and\
  \citenamefont {Pichler}}]{Kramberger2008}%
  \BibitemOpen
  \bibfield  {author} {\bibinfo {author} {\bibfnamefont {C.}~\bibnamefont
  {Kramberger}}, \bibinfo {author} {\bibfnamefont {R.}~\bibnamefont {Hambach}},
  \bibinfo {author} {\bibfnamefont {C.}~\bibnamefont {Giorgetti}}, \bibinfo
  {author} {\bibfnamefont {M.~H.}\ \bibnamefont {R\"ummeli}}, \bibinfo {author}
  {\bibfnamefont {M.}~\bibnamefont {Knupfer}}, \bibinfo {author} {\bibfnamefont
  {J.}~\bibnamefont {Fink}}, \bibinfo {author} {\bibfnamefont {B.}~\bibnamefont
  {B\"uchner}}, \bibinfo {author} {\bibfnamefont {L.}~\bibnamefont {Reining}},
  \bibinfo {author} {\bibfnamefont {E.}~\bibnamefont {Einarsson}}, \bibinfo
  {author} {\bibfnamefont {S.}~\bibnamefont {Maruyama}}, \bibinfo {author}
  {\bibfnamefont {F.}~\bibnamefont {Sottile}}, \bibinfo {author} {\bibfnamefont
  {K.}~\bibnamefont {Hannewald}}, \bibinfo {author} {\bibfnamefont
  {V.}~\bibnamefont {Olevano}}, \bibinfo {author} {\bibfnamefont {A.~G.}\
  \bibnamefont {Marinopoulos}}, \ and\ \bibinfo {author} {\bibfnamefont
  {T.}~\bibnamefont {Pichler}},\ }\href {\doibase
  10.1103/PhysRevLett.100.196803} {\bibfield  {journal} {\bibinfo  {journal}
  {Phys. Rev. Lett.}\ }\textbf {\bibinfo {volume} {100}},\ \bibinfo {pages}
  {196803} (\bibinfo {year} {2008})}\BibitemShut {NoStop}%
\bibitem [{\citenamefont {Roth}\ \emph
  {et~al.}(2011{\natexlab{a}})\citenamefont {Roth}, \citenamefont {Mahns},
  \citenamefont {B\"uchner},\ and\ \citenamefont {Knupfer}}]{Roth_2011_b}%
  \BibitemOpen
  \bibfield  {author} {\bibinfo {author} {\bibfnamefont {F.}~\bibnamefont
  {Roth}}, \bibinfo {author} {\bibfnamefont {B.}~\bibnamefont {Mahns}},
  \bibinfo {author} {\bibfnamefont {B.}~\bibnamefont {B\"uchner}}, \ and\
  \bibinfo {author} {\bibfnamefont {M.}~\bibnamefont {Knupfer}},\ }\href
  {\doibase 10.1103/PhysRevB.83.165436} {\bibfield  {journal} {\bibinfo
  {journal} {Phys. Rev. B}\ }\textbf {\bibinfo {volume} {83}},\ \bibinfo
  {pages} {165436} (\bibinfo {year} {2011}{\natexlab{a}})}\BibitemShut
  {NoStop}%
\bibitem [{\citenamefont {Crecelius}\ \emph {et~al.}(1983)\citenamefont
  {Crecelius}, \citenamefont {Fink}, \citenamefont {Ritsko}, \citenamefont
  {Stamm}, \citenamefont {Freund},\ and\ \citenamefont
  {Gonska}}]{Crecelius1983}%
  \BibitemOpen
  \bibfield  {author} {\bibinfo {author} {\bibfnamefont {G.}~\bibnamefont
  {Crecelius}}, \bibinfo {author} {\bibfnamefont {J.}~\bibnamefont {Fink}},
  \bibinfo {author} {\bibfnamefont {J.~J.}\ \bibnamefont {Ritsko}}, \bibinfo
  {author} {\bibfnamefont {M.}~\bibnamefont {Stamm}}, \bibinfo {author}
  {\bibfnamefont {H.-J.}\ \bibnamefont {Freund}}, \ and\ \bibinfo {author}
  {\bibfnamefont {H.}~\bibnamefont {Gonska}},\ }\href {\doibase
  10.1103/PhysRevB.28.1802} {\bibfield  {journal} {\bibinfo  {journal} {Phys.
  Rev. B}\ }\textbf {\bibinfo {volume} {28}},\ \bibinfo {pages} {1802}
  (\bibinfo {year} {1983})}\BibitemShut {NoStop}%
\bibitem [{\citenamefont {Ritsko}\ \emph {et~al.}(1983)\citenamefont {Ritsko},
  \citenamefont {Crecelius},\ and\ \citenamefont {Fink}}]{Ritsko1983}%
  \BibitemOpen
  \bibfield  {author} {\bibinfo {author} {\bibfnamefont {J.~J.}\ \bibnamefont
  {Ritsko}}, \bibinfo {author} {\bibfnamefont {G.}~\bibnamefont {Crecelius}}, \
  and\ \bibinfo {author} {\bibfnamefont {J.}~\bibnamefont {Fink}},\ }\href
  {\doibase 10.1103/PhysRevB.27.4902} {\bibfield  {journal} {\bibinfo
  {journal} {Phys. Rev. B}\ }\textbf {\bibinfo {volume} {27}},\ \bibinfo
  {pages} {4902} (\bibinfo {year} {1983})}\BibitemShut {NoStop}%
\bibitem [{\citenamefont {Pellegrin}\ \emph {et~al.}(1991)\citenamefont
  {Pellegrin}, \citenamefont {Fink},\ and\ \citenamefont
  {Drechsler}}]{Pellegrin1991}%
  \BibitemOpen
  \bibfield  {author} {\bibinfo {author} {\bibfnamefont {E.}~\bibnamefont
  {Pellegrin}}, \bibinfo {author} {\bibfnamefont {J.}~\bibnamefont {Fink}}, \
  and\ \bibinfo {author} {\bibfnamefont {S.~L.}\ \bibnamefont {Drechsler}},\
  }\href {\doibase 10.1103/PhysRevLett.66.2022} {\bibfield  {journal} {\bibinfo
   {journal} {Phys. Rev. Lett.}\ }\textbf {\bibinfo {volume} {66}},\ \bibinfo
  {pages} {2022} (\bibinfo {year} {1991})}\BibitemShut {NoStop}%
\bibitem [{\citenamefont {Fink}\ and\ \citenamefont
  {Leising}(1986)}]{FinkLeising1986}%
  \BibitemOpen
  \bibfield  {author} {\bibinfo {author} {\bibfnamefont {J.}~\bibnamefont
  {Fink}}\ and\ \bibinfo {author} {\bibfnamefont {G.}~\bibnamefont {Leising}},\
  }\href {\doibase 10.1103/PhysRevB.34.5320} {\bibfield  {journal} {\bibinfo
  {journal} {Phys. Rev. B}\ }\textbf {\bibinfo {volume} {34}},\ \bibinfo
  {pages} {5320} (\bibinfo {year} {1986})}\BibitemShut {NoStop}%
\bibitem [{\citenamefont {Marinopoulos}\ \emph {et~al.}(2002)\citenamefont
  {Marinopoulos}, \citenamefont {Reining}, \citenamefont {Olevano},
  \citenamefont {Rubio}, \citenamefont {Pichler}, \citenamefont {Liu},
  \citenamefont {Knupfer},\ and\ \citenamefont {Fink}}]{Marinopoulos2002}%
  \BibitemOpen
  \bibfield  {author} {\bibinfo {author} {\bibfnamefont {A.~G.}\ \bibnamefont
  {Marinopoulos}}, \bibinfo {author} {\bibfnamefont {L.}~\bibnamefont
  {Reining}}, \bibinfo {author} {\bibfnamefont {V.}~\bibnamefont {Olevano}},
  \bibinfo {author} {\bibfnamefont {A.}~\bibnamefont {Rubio}}, \bibinfo
  {author} {\bibfnamefont {T.}~\bibnamefont {Pichler}}, \bibinfo {author}
  {\bibfnamefont {X.}~\bibnamefont {Liu}}, \bibinfo {author} {\bibfnamefont
  {M.}~\bibnamefont {Knupfer}}, \ and\ \bibinfo {author} {\bibfnamefont
  {J.}~\bibnamefont {Fink}},\ }\href {\doibase 10.1103/PhysRevLett.89.076402}
  {\bibfield  {journal} {\bibinfo  {journal} {Phys. Rev. Lett.}\ }\textbf
  {\bibinfo {volume} {89}},\ \bibinfo {pages} {076402} (\bibinfo {year}
  {2002})}\BibitemShut {NoStop}%
\bibitem [{\citenamefont {Waidmann}\ \emph {et~al.}(2000)\citenamefont
  {Waidmann}, \citenamefont {Knupfer}, \citenamefont {Arnold}, \citenamefont
  {Fink}, \citenamefont {Fleszar},\ and\ \citenamefont {Hanke}}]{Waidmann2000}%
  \BibitemOpen
  \bibfield  {author} {\bibinfo {author} {\bibfnamefont {S.}~\bibnamefont
  {Waidmann}}, \bibinfo {author} {\bibfnamefont {M.}~\bibnamefont {Knupfer}},
  \bibinfo {author} {\bibfnamefont {B.}~\bibnamefont {Arnold}}, \bibinfo
  {author} {\bibfnamefont {J.}~\bibnamefont {Fink}}, \bibinfo {author}
  {\bibfnamefont {A.}~\bibnamefont {Fleszar}}, \ and\ \bibinfo {author}
  {\bibfnamefont {W.}~\bibnamefont {Hanke}},\ }\href {\doibase
  10.1103/PhysRevB.61.10149} {\bibfield  {journal} {\bibinfo  {journal} {Phys.
  Rev. B}\ }\textbf {\bibinfo {volume} {61}},\ \bibinfo {pages} {10149}
  (\bibinfo {year} {2000})}\BibitemShut {NoStop}%
\bibitem [{\citenamefont {Aryasetiawan}\ \emph {et~al.}(1994)\citenamefont
  {Aryasetiawan}, \citenamefont {Gunnarsson}, \citenamefont {Knupfer},\ and\
  \citenamefont {Fink}}]{Aryasetiawan1994}%
  \BibitemOpen
  \bibfield  {author} {\bibinfo {author} {\bibfnamefont {F.}~\bibnamefont
  {Aryasetiawan}}, \bibinfo {author} {\bibfnamefont {O.}~\bibnamefont
  {Gunnarsson}}, \bibinfo {author} {\bibfnamefont {M.}~\bibnamefont {Knupfer}},
  \ and\ \bibinfo {author} {\bibfnamefont {J.}~\bibnamefont {Fink}},\ }\href
  {\doibase 10.1103/PhysRevB.50.7311} {\bibfield  {journal} {\bibinfo
  {journal} {Phys. Rev. B}\ }\textbf {\bibinfo {volume} {50}},\ \bibinfo
  {pages} {7311} (\bibinfo {year} {1994})}\BibitemShut {NoStop}%
\bibitem [{\citenamefont {Van~Hove}(1954)}]{Hove1954}%
  \BibitemOpen
  \bibfield  {author} {\bibinfo {author} {\bibfnamefont {L.}~\bibnamefont
  {Van~Hove}},\ }\href {\doibase 10.1103/PhysRev.95.249} {\bibfield  {journal}
  {\bibinfo  {journal} {Phys. Rev.}\ }\textbf {\bibinfo {volume} {95}},\
  \bibinfo {pages} {249} (\bibinfo {year} {1954})}\BibitemShut {NoStop}%
\bibitem [{\citenamefont {Mahan}(2000)}]{Mahan2000}%
  \BibitemOpen
  \bibfield  {author} {\bibinfo {author} {\bibfnamefont {G.~D.}\ \bibnamefont
  {Mahan}},\ }\href@noop {} {\emph {\bibinfo {title} {Many Particle Physics
  (Physics of Solids and Liquids)}}},\ \bibinfo {edition} {3rd}\ ed.\ (\bibinfo
   {publisher} {Springer},\ \bibinfo {year} {2000})\BibitemShut {NoStop}%
\bibitem [{\citenamefont {Kubo}(1957)}]{Kubo1957}%
  \BibitemOpen
  \bibfield  {author} {\bibinfo {author} {\bibfnamefont {R.}~\bibnamefont
  {Kubo}},\ }\href {\doibase 10.1143/JPSJ.12.570} {\bibfield  {journal}
  {\bibinfo  {journal} {J. Phys. Soc. Jpn.}\ }\textbf {\bibinfo {volume}
  {12}},\ \bibinfo {pages} {570} (\bibinfo {year} {1957})}\BibitemShut
  {NoStop}%
\bibitem [{\citenamefont {Hanke}(1978)}]{HANKE1978}%
  \BibitemOpen
  \bibfield  {author} {\bibinfo {author} {\bibfnamefont {W.}~\bibnamefont
  {Hanke}},\ }\href {\doibase 10.1080/00018737800101384} {\bibfield  {journal}
  {\bibinfo  {journal} {Adv. Phys.}\ }\textbf {\bibinfo {volume} {27}},\
  \bibinfo {pages} {287} (\bibinfo {year} {1978})}\BibitemShut {NoStop}%
\bibitem [{\citenamefont {Nyquist}(1928)}]{Nyquist1928}%
  \BibitemOpen
  \bibfield  {author} {\bibinfo {author} {\bibfnamefont {H.}~\bibnamefont
  {Nyquist}},\ }\href {\doibase 10.1103/PhysRev.32.110} {\bibfield  {journal}
  {\bibinfo  {journal} {Phys. Rev.}\ }\textbf {\bibinfo {volume} {32}},\
  \bibinfo {pages} {110} (\bibinfo {year} {1928})}\BibitemShut {NoStop}%
\bibitem [{\citenamefont {Callen}\ and\ \citenamefont
  {Welton}(1951)}]{Callen1951}%
  \BibitemOpen
  \bibfield  {author} {\bibinfo {author} {\bibfnamefont {H.~B.}\ \bibnamefont
  {Callen}}\ and\ \bibinfo {author} {\bibfnamefont {T.~A.}\ \bibnamefont
  {Welton}},\ }\href {\doibase 10.1103/PhysRev.83.34} {\bibfield  {journal}
  {\bibinfo  {journal} {Phys. Rev.}\ }\textbf {\bibinfo {volume} {83}},\
  \bibinfo {pages} {34} (\bibinfo {year} {1951})}\BibitemShut {NoStop}%
\bibitem [{\citenamefont {Dressel}\ and\ \citenamefont
  {Gr\"uner}(2002)}]{Dressel2002}%
  \BibitemOpen
  \bibfield  {author} {\bibinfo {author} {\bibfnamefont {M.}~\bibnamefont
  {Dressel}}\ and\ \bibinfo {author} {\bibfnamefont {G.}~\bibnamefont
  {Gr\"uner}},\ }\href@noop {} {\emph {\bibinfo {title} {Electrodynamics of
  Solids: Optical Properties of Electrons in Matter}}},\ \bibinfo {edition}
  {1st}\ ed.\ (\bibinfo  {publisher} {Cambridge University Press},\ \bibinfo
  {year} {2002})\BibitemShut {NoStop}%
\bibitem [{\citenamefont {Gross}\ and\ \citenamefont {Marx}(2012)}]{Gross2012}%
  \BibitemOpen
  \bibfield  {author} {\bibinfo {author} {\bibfnamefont {R.}~\bibnamefont
  {Gross}}\ and\ \bibinfo {author} {\bibfnamefont {A.}~\bibnamefont {Marx}},\
  }\href@noop {} {\emph {\bibinfo {title} {Festkörperphysik}}}\ (\bibinfo
  {publisher} {Oldenbourg Wissenschaftsverlag},\ \bibinfo {year}
  {2012})\BibitemShut {NoStop}%
\bibitem [{\citenamefont {Fuggle}\ and\ \citenamefont
  {Inglesfield}(1992)}]{Fuggle1992}%
  \BibitemOpen
  \bibinfo {editor} {\bibfnamefont {J.~C.}\ \bibnamefont {Fuggle}}\ and\
  \bibinfo {editor} {\bibfnamefont {J.~E.}\ \bibnamefont {Inglesfield}},\
  eds.,\ \href@noop {} {\emph {\bibinfo {title} {Unoccupied Electronic States
  (Topics in Applied Physics vol. 69)}}}\ (\bibinfo  {publisher} {Springer,
  Berlin},\ \bibinfo {year} {1992})\BibitemShut {NoStop}%
\bibitem [{\citenamefont {Fink}\ and\ \citenamefont {Kisker}(1980)}]{Fink1980}%
  \BibitemOpen
  \bibfield  {author} {\bibinfo {author} {\bibfnamefont {J.}~\bibnamefont
  {Fink}}\ and\ \bibinfo {author} {\bibfnamefont {E.}~\bibnamefont {Kisker}},\
  }\href {\doibase http://dx.doi.org/10.1063/1.1136338} {\bibfield  {journal}
  {\bibinfo  {journal} {Rev. Sci. Instrum.}\ }\textbf {\bibinfo {volume}
  {51}},\ \bibinfo {pages} {918} (\bibinfo {year} {1980})}\BibitemShut
  {NoStop}%
\bibitem [{\citenamefont {Damascelli}\ \emph {et~al.}(2003)\citenamefont
  {Damascelli}, \citenamefont {Hussain},\ and\ \citenamefont
  {Shen}}]{Damascelli2003}%
  \BibitemOpen
  \bibfield  {author} {\bibinfo {author} {\bibfnamefont {A.}~\bibnamefont
  {Damascelli}}, \bibinfo {author} {\bibfnamefont {Z.}~\bibnamefont {Hussain}},
  \ and\ \bibinfo {author} {\bibfnamefont {Z.-X.}\ \bibnamefont {Shen}},\
  }\href {\doibase 10.1103/RevModPhys.75.473} {\bibfield  {journal} {\bibinfo
  {journal} {Rev. Mod. Phys.}\ }\textbf {\bibinfo {volume} {75}},\ \bibinfo
  {pages} {473} (\bibinfo {year} {2003})}\BibitemShut {NoStop}%
\bibitem [{\citenamefont {Roth}\ \emph {et~al.}(2012)\citenamefont {Roth},
  \citenamefont {Schuster}, \citenamefont {K\"onig}, \citenamefont {Knupfer},\
  and\ \citenamefont {Berger}}]{Roth_penta}%
  \BibitemOpen
  \bibfield  {author} {\bibinfo {author} {\bibfnamefont {F.}~\bibnamefont
  {Roth}}, \bibinfo {author} {\bibfnamefont {R.}~\bibnamefont {Schuster}},
  \bibinfo {author} {\bibfnamefont {A.}~\bibnamefont {K\"onig}}, \bibinfo
  {author} {\bibfnamefont {M.}~\bibnamefont {Knupfer}}, \ and\ \bibinfo
  {author} {\bibfnamefont {H.}~\bibnamefont {Berger}},\ }\href {\doibase
  10.1063/1.4723812} {\bibfield  {journal} {\bibinfo  {journal} {J. Chem.
  Phys.}\ }\textbf {\bibinfo {volume} {136}},\ \bibinfo {pages} {204708}
  (\bibinfo {year} {2012})}\BibitemShut {NoStop}%
\bibitem [{\citenamefont {Egerton}\ and\ \citenamefont
  {Cheng}(1987)}]{Egerton1987}%
  \BibitemOpen
  \bibfield  {author} {\bibinfo {author} {\bibfnamefont {R.}~\bibnamefont
  {Egerton}}\ and\ \bibinfo {author} {\bibfnamefont {S.}~\bibnamefont
  {Cheng}},\ }\href {\doibase http://dx.doi.org/10.1016/0304-3991(87)90148-3}
  {\bibfield  {journal} {\bibinfo  {journal} {Ultramicroscopy}\ }\textbf
  {\bibinfo {volume} {21}},\ \bibinfo {pages} {231 } (\bibinfo {year}
  {1987})}\BibitemShut {NoStop}%
\bibitem [{\citenamefont {Wang}\ \emph {et~al.}(2010)\citenamefont {Wang},
  \citenamefont {Zhang}, \citenamefont {Cao}, \citenamefont {Nishi},\ and\
  \citenamefont {Takaoka}}]{Wang2010}%
  \BibitemOpen
  \bibfield  {author} {\bibinfo {author} {\bibfnamefont {F.}~\bibnamefont
  {Wang}}, \bibinfo {author} {\bibfnamefont {H.-B.}\ \bibnamefont {Zhang}},
  \bibinfo {author} {\bibfnamefont {M.}~\bibnamefont {Cao}}, \bibinfo {author}
  {\bibfnamefont {R.}~\bibnamefont {Nishi}}, \ and\ \bibinfo {author}
  {\bibfnamefont {A.}~\bibnamefont {Takaoka}},\ }\href {\doibase
  http://dx.doi.org/10.1016/j.micron.2010.05.014} {\bibfield  {journal}
  {\bibinfo  {journal} {Micron}\ }\textbf {\bibinfo {volume} {41}},\ \bibinfo
  {pages} {769 } (\bibinfo {year} {2010})}\BibitemShut {NoStop}%
\bibitem [{\citenamefont {Pozsgai}(2007)}]{Pozsgai2007}%
  \BibitemOpen
  \bibfield  {author} {\bibinfo {author} {\bibfnamefont {I.}~\bibnamefont
  {Pozsgai}},\ }\href {\doibase
  http://dx.doi.org/10.1016/j.ultramic.2006.07.005} {\bibfield  {journal}
  {\bibinfo  {journal} {Ultramicroscopy}\ }\textbf {\bibinfo {volume} {107}},\
  \bibinfo {pages} {191 } (\bibinfo {year} {2007})}\BibitemShut {NoStop}%
\bibitem [{\citenamefont {Zhang}\ \emph {et~al.}(2012)\citenamefont {Zhang},
  \citenamefont {Egerton},\ and\ \citenamefont {Malac}}]{Zhang2012}%
  \BibitemOpen
  \bibfield  {author} {\bibinfo {author} {\bibfnamefont {H.-R.}\ \bibnamefont
  {Zhang}}, \bibinfo {author} {\bibfnamefont {R.~F.}\ \bibnamefont {Egerton}},
  \ and\ \bibinfo {author} {\bibfnamefont {M.}~\bibnamefont {Malac}},\ }\href
  {\doibase http://dx.doi.org/10.1016/j.micron.2011.07.003} {\bibfield
  {journal} {\bibinfo  {journal} {Micron}\ }\textbf {\bibinfo {volume} {43}},\
  \bibinfo {pages} {8 } (\bibinfo {year} {2012})}\BibitemShut {NoStop}%
\bibitem [{\citenamefont {Roth}\ \emph {et~al.}(2013)\citenamefont {Roth},
  \citenamefont {K\"onig}, \citenamefont {Kramberger}, \citenamefont {Pichler},
  \citenamefont {B\"uchner},\ and\ \citenamefont {Knupfer}}]{Roth_HOPG}%
  \BibitemOpen
  \bibfield  {author} {\bibinfo {author} {\bibfnamefont {F.}~\bibnamefont
  {Roth}}, \bibinfo {author} {\bibfnamefont {A.}~\bibnamefont {K\"onig}},
  \bibinfo {author} {\bibfnamefont {C.}~\bibnamefont {Kramberger}}, \bibinfo
  {author} {\bibfnamefont {T.}~\bibnamefont {Pichler}}, \bibinfo {author}
  {\bibfnamefont {B.}~\bibnamefont {B\"uchner}}, \ and\ \bibinfo {author}
  {\bibfnamefont {M.}~\bibnamefont {Knupfer}},\ }\href
  {http://stacks.iop.org/0295-5075/102/i=1/a=17001} {\bibfield  {journal}
  {\bibinfo  {journal} {Europhys. Lett.}\ }\textbf {\bibinfo {volume} {102}},\
  \bibinfo {pages} {17001} (\bibinfo {year} {2013})}\BibitemShut {NoStop}%
\bibitem [{\citenamefont {Geim}(2009)}]{Geim2009}%
  \BibitemOpen
  \bibfield  {author} {\bibinfo {author} {\bibfnamefont {A.~K.}\ \bibnamefont
  {Geim}},\ }\href {\doibase 10.1126/science.1158877} {\bibfield  {journal}
  {\bibinfo  {journal} {Science}\ }\textbf {\bibinfo {volume} {324}},\ \bibinfo
  {pages} {1530} (\bibinfo {year} {2009})}\BibitemShut {NoStop}%
\bibitem [{\citenamefont {Pines}(1956)}]{Pines1956}%
  \BibitemOpen
  \bibfield  {author} {\bibinfo {author} {\bibfnamefont {D.}~\bibnamefont
  {Pines}},\ }\href {\doibase 10.1103/RevModPhys.28.184} {\bibfield  {journal}
  {\bibinfo  {journal} {Rev. Mod. Phys.}\ }\textbf {\bibinfo {volume} {28}},\
  \bibinfo {pages} {184} (\bibinfo {year} {1956})}\BibitemShut {NoStop}%
\bibitem [{\citenamefont {Raimes}(1957)}]{Raimes1957}%
  \BibitemOpen
  \bibfield  {author} {\bibinfo {author} {\bibfnamefont {S.}~\bibnamefont
  {Raimes}},\ }\href {http://stacks.iop.org/0034-4885/20/i=1/a=301} {\bibfield
  {journal} {\bibinfo  {journal} {Rep. Prog. Phys.}\ }\textbf {\bibinfo
  {volume} {20}},\ \bibinfo {pages} {1} (\bibinfo {year} {1957})}\BibitemShut
  {NoStop}%
\bibitem [{\citenamefont {Lindhard}(1954)}]{LINDHARD1954}%
  \BibitemOpen
  \bibfield  {author} {\bibinfo {author} {\bibfnamefont {J.}~\bibnamefont
  {Lindhard}},\ }\href@noop {} {\bibfield  {journal} {\bibinfo  {journal} {Kgl.
  Danske Videnskab.Selskab, Mat.-fys. Medd.}\ }\textbf {\bibinfo {volume}
  {28}},\ \bibinfo {pages} {1} (\bibinfo {year} {1954})}\BibitemShut {NoStop}%
\bibitem [{\citenamefont {Nolting}(2009)}]{Nolting2009}%
  \BibitemOpen
  \bibfield  {author} {\bibinfo {author} {\bibfnamefont {W.}~\bibnamefont
  {Nolting}},\ }\href@noop {} {\emph {\bibinfo {title} {Fundamentals of
  Many-body Physics}}}\ (\bibinfo  {publisher} {Springer},\ \bibinfo {year}
  {2009})\BibitemShut {NoStop}%
\bibitem [{\citenamefont {Gibbons}\ \emph {et~al.}(1976)\citenamefont
  {Gibbons}, \citenamefont {Schnatterly}, \citenamefont {Ritsko},\ and\
  \citenamefont {Fields}}]{Gibbons1976}%
  \BibitemOpen
  \bibfield  {author} {\bibinfo {author} {\bibfnamefont {P.~C.}\ \bibnamefont
  {Gibbons}}, \bibinfo {author} {\bibfnamefont {S.~E.}\ \bibnamefont
  {Schnatterly}}, \bibinfo {author} {\bibfnamefont {J.~J.}\ \bibnamefont
  {Ritsko}}, \ and\ \bibinfo {author} {\bibfnamefont {J.~R.}\ \bibnamefont
  {Fields}},\ }\href {\doibase 10.1103/PhysRevB.13.2451} {\bibfield  {journal}
  {\bibinfo  {journal} {Phys. Rev. B}\ }\textbf {\bibinfo {volume} {13}},\
  \bibinfo {pages} {2451} (\bibinfo {year} {1976})}\BibitemShut {NoStop}%
\bibitem [{\citenamefont {N\"ucker}\ \emph {et~al.}(1989)\citenamefont
  {N\"ucker}, \citenamefont {Romberg}, \citenamefont {Nakai}, \citenamefont
  {Scheerer}, \citenamefont {Fink}, \citenamefont {Yan},\ and\ \citenamefont
  {Zhao}}]{Nucker1989}%
  \BibitemOpen
  \bibfield  {author} {\bibinfo {author} {\bibfnamefont {N.}~\bibnamefont
  {N\"ucker}}, \bibinfo {author} {\bibfnamefont {H.}~\bibnamefont {Romberg}},
  \bibinfo {author} {\bibfnamefont {S.}~\bibnamefont {Nakai}}, \bibinfo
  {author} {\bibfnamefont {B.}~\bibnamefont {Scheerer}}, \bibinfo {author}
  {\bibfnamefont {J.}~\bibnamefont {Fink}}, \bibinfo {author} {\bibfnamefont
  {Y.~F.}\ \bibnamefont {Yan}}, \ and\ \bibinfo {author} {\bibfnamefont
  {Z.~X.}\ \bibnamefont {Zhao}},\ }\href {\doibase 10.1103/PhysRevB.39.12379}
  {\bibfield  {journal} {\bibinfo  {journal} {Phys. Rev. B}\ }\textbf {\bibinfo
  {volume} {39}},\ \bibinfo {pages} {12379} (\bibinfo {year}
  {1989})}\BibitemShut {NoStop}%
\bibitem [{\citenamefont {Mitsuhashi}\ \emph {et~al.}(2010)\citenamefont
  {Mitsuhashi}, \citenamefont {Suzuki}, \citenamefont {Yamanari}, \citenamefont
  {Mitamura}, \citenamefont {Kambe}, \citenamefont {Ikeda}, \citenamefont
  {Okamoto}, \citenamefont {Fujiwara}, \citenamefont {Yamaji}, \citenamefont
  {Kawasaki}, \citenamefont {Maniwa},\ and\ \citenamefont
  {Kubozono}}]{Mitsuhashi2010}%
  \BibitemOpen
  \bibfield  {author} {\bibinfo {author} {\bibfnamefont {R.}~\bibnamefont
  {Mitsuhashi}}, \bibinfo {author} {\bibfnamefont {Y.}~\bibnamefont {Suzuki}},
  \bibinfo {author} {\bibfnamefont {Y.}~\bibnamefont {Yamanari}}, \bibinfo
  {author} {\bibfnamefont {H.}~\bibnamefont {Mitamura}}, \bibinfo {author}
  {\bibfnamefont {T.}~\bibnamefont {Kambe}}, \bibinfo {author} {\bibfnamefont
  {N.}~\bibnamefont {Ikeda}}, \bibinfo {author} {\bibfnamefont
  {H.}~\bibnamefont {Okamoto}}, \bibinfo {author} {\bibfnamefont
  {A.}~\bibnamefont {Fujiwara}}, \bibinfo {author} {\bibfnamefont
  {M.}~\bibnamefont {Yamaji}}, \bibinfo {author} {\bibfnamefont
  {N.}~\bibnamefont {Kawasaki}}, \bibinfo {author} {\bibfnamefont
  {Y.}~\bibnamefont {Maniwa}}, \ and\ \bibinfo {author} {\bibfnamefont
  {Y.}~\bibnamefont {Kubozono}},\ }\href@noop {} {\bibfield  {journal}
  {\bibinfo  {journal} {Nature}\ }\textbf {\bibinfo {volume} {464}},\ \bibinfo
  {pages} {76} (\bibinfo {year} {2010})}\BibitemShut {NoStop}%
\bibitem [{\citenamefont {De}\ \emph {et~al.}(1985)\citenamefont {De},
  \citenamefont {Ghosh}, \citenamefont {Roychowdhury},\ and\ \citenamefont
  {Roychowdhury}}]{De1985}%
  \BibitemOpen
  \bibfield  {author} {\bibinfo {author} {\bibfnamefont {A.}~\bibnamefont
  {De}}, \bibinfo {author} {\bibfnamefont {R.}~\bibnamefont {Ghosh}}, \bibinfo
  {author} {\bibfnamefont {S.}~\bibnamefont {Roychowdhury}}, \ and\ \bibinfo
  {author} {\bibfnamefont {P.}~\bibnamefont {Roychowdhury}},\ }\href {\doibase
  10.1107/S0108270185005959} {\bibfield  {journal} {\bibinfo  {journal} {Acta
  Crystallogr. Sec. C}\ }\textbf {\bibinfo {volume} {41}},\ \bibinfo {pages}
  {907} (\bibinfo {year} {1985})}\BibitemShut {NoStop}%
\bibitem [{\citenamefont {Hebard}\ \emph {et~al.}(1991)\citenamefont {Hebard},
  \citenamefont {Rosseinsky}, \citenamefont {Haddon}, \citenamefont {Murphy},
  \citenamefont {Glarum}, \citenamefont {Palastra}, \citenamefont {Ramirez},\
  and\ \citenamefont {Kortan}}]{Hebard1991}%
  \BibitemOpen
  \bibfield  {author} {\bibinfo {author} {\bibfnamefont {A.}~\bibnamefont
  {Hebard}}, \bibinfo {author} {\bibfnamefont {M.}~\bibnamefont {Rosseinsky}},
  \bibinfo {author} {\bibfnamefont {R.}~\bibnamefont {Haddon}}, \bibinfo
  {author} {\bibfnamefont {D.}~\bibnamefont {Murphy}}, \bibinfo {author}
  {\bibfnamefont {S.}~\bibnamefont {Glarum}}, \bibinfo {author} {\bibfnamefont
  {T.}~\bibnamefont {Palastra}}, \bibinfo {author} {\bibfnamefont
  {A.}~\bibnamefont {Ramirez}}, \ and\ \bibinfo {author} {\bibfnamefont
  {A.}~\bibnamefont {Kortan}},\ }\href@noop {} {\bibfield  {journal} {\bibinfo
  {journal} {Nature}\ }\textbf {\bibinfo {volume} {350}},\ \bibinfo {pages}
  {600} (\bibinfo {year} {1991})}\BibitemShut {NoStop}%
\bibitem [{\citenamefont {Gunnarson}(2004)}]{Gunnarsson2004}%
  \BibitemOpen
  \bibfield  {author} {\bibinfo {author} {\bibfnamefont {O.}~\bibnamefont
  {Gunnarson}},\ }\href@noop {} {\emph {\bibinfo {title} {Alkali Doped
  Fullerides}}}\ (\bibinfo  {publisher} {World Scientific, Singapore},\
  \bibinfo {year} {2004})\BibitemShut {NoStop}%
\bibitem [{\citenamefont {Roth}\ \emph
  {et~al.}(2011{\natexlab{b}})\citenamefont {Roth}, \citenamefont {Mahns},
  \citenamefont {B\"uchner},\ and\ \citenamefont {Knupfer}}]{Roth2011_c}%
  \BibitemOpen
  \bibfield  {author} {\bibinfo {author} {\bibfnamefont {F.}~\bibnamefont
  {Roth}}, \bibinfo {author} {\bibfnamefont {B.}~\bibnamefont {Mahns}},
  \bibinfo {author} {\bibfnamefont {B.}~\bibnamefont {B\"uchner}}, \ and\
  \bibinfo {author} {\bibfnamefont {M.}~\bibnamefont {Knupfer}},\ }\href
  {\doibase 10.1103/PhysRevB.83.144501} {\bibfield  {journal} {\bibinfo
  {journal} {Phys. Rev. B}\ }\textbf {\bibinfo {volume} {83}},\ \bibinfo
  {pages} {144501} (\bibinfo {year} {2011}{\natexlab{b}})}\BibitemShut
  {NoStop}%
\bibitem [{\citenamefont {Roth}\ \emph {et~al.}(2010)\citenamefont {Roth},
  \citenamefont {Gatti}, \citenamefont {Cudazzo}, \citenamefont {Grobosch},
  \citenamefont {Mahns}, \citenamefont {B\"uchner}, \citenamefont {Rubio},\
  and\ \citenamefont {Knupfer}}]{Roth2010}%
  \BibitemOpen
  \bibfield  {author} {\bibinfo {author} {\bibfnamefont {F.}~\bibnamefont
  {Roth}}, \bibinfo {author} {\bibfnamefont {M.}~\bibnamefont {Gatti}},
  \bibinfo {author} {\bibfnamefont {P.}~\bibnamefont {Cudazzo}}, \bibinfo
  {author} {\bibfnamefont {M.}~\bibnamefont {Grobosch}}, \bibinfo {author}
  {\bibfnamefont {B.}~\bibnamefont {Mahns}}, \bibinfo {author} {\bibfnamefont
  {B.}~\bibnamefont {B\"uchner}}, \bibinfo {author} {\bibfnamefont
  {A.}~\bibnamefont {Rubio}}, \ and\ \bibinfo {author} {\bibfnamefont
  {M.}~\bibnamefont {Knupfer}},\ }\href@noop {} {\bibfield  {journal} {\bibinfo
   {journal} {New J. Phys.}\ }\textbf {\bibinfo {volume} {12}},\ \bibinfo
  {pages} {103036} (\bibinfo {year} {2010})}\BibitemShut {NoStop}%
\bibitem [{\citenamefont {Roth}\ \emph
  {et~al.}(2011{\natexlab{c}})\citenamefont {Roth}, \citenamefont {Mahns},
  \citenamefont {B\"uchner},\ and\ \citenamefont {Knupfer}}]{Roth2011}%
  \BibitemOpen
  \bibfield  {author} {\bibinfo {author} {\bibfnamefont {F.}~\bibnamefont
  {Roth}}, \bibinfo {author} {\bibfnamefont {B.}~\bibnamefont {Mahns}},
  \bibinfo {author} {\bibfnamefont {B.}~\bibnamefont {B\"uchner}}, \ and\
  \bibinfo {author} {\bibfnamefont {M.}~\bibnamefont {Knupfer}},\ }\href
  {\doibase 10.1103/PhysRevB.83.144501} {\bibfield  {journal} {\bibinfo
  {journal} {Phys. Rev. B}\ }\textbf {\bibinfo {volume} {83}},\ \bibinfo
  {pages} {144501} (\bibinfo {year} {2011}{\natexlab{c}})}\BibitemShut
  {NoStop}%
\bibitem [{\citenamefont {Gunnarsson}\ \emph
  {et~al.}(1996{\natexlab{a}})\citenamefont {Gunnarsson}, \citenamefont
  {Liechtenstein}, \citenamefont {Eyert}, \citenamefont {Knupfer},
  \citenamefont {Fink},\ and\ \citenamefont {Armbruster}}]{Gunnarsson1996}%
  \BibitemOpen
  \bibfield  {author} {\bibinfo {author} {\bibfnamefont {O.}~\bibnamefont
  {Gunnarsson}}, \bibinfo {author} {\bibfnamefont {A.~I.}\ \bibnamefont
  {Liechtenstein}}, \bibinfo {author} {\bibfnamefont {V.}~\bibnamefont
  {Eyert}}, \bibinfo {author} {\bibfnamefont {M.}~\bibnamefont {Knupfer}},
  \bibinfo {author} {\bibfnamefont {J.}~\bibnamefont {Fink}}, \ and\ \bibinfo
  {author} {\bibfnamefont {J.~F.}\ \bibnamefont {Armbruster}},\ }\href
  {\doibase 10.1103/PhysRevB.53.3455} {\bibfield  {journal} {\bibinfo
  {journal} {Phys. Rev. B}\ }\textbf {\bibinfo {volume} {53}},\ \bibinfo
  {pages} {3455} (\bibinfo {year} {1996}{\natexlab{a}})}\BibitemShut {NoStop}%
\bibitem [{\citenamefont {Gunnarsson}\ \emph
  {et~al.}(1996{\natexlab{b}})\citenamefont {Gunnarsson}, \citenamefont
  {Eyert}, \citenamefont {Knupfer}, \citenamefont {Fink},\ and\ \citenamefont
  {Armbruster}}]{Gunnarsson1996_b}%
  \BibitemOpen
  \bibfield  {author} {\bibinfo {author} {\bibfnamefont {O.}~\bibnamefont
  {Gunnarsson}}, \bibinfo {author} {\bibfnamefont {V.}~\bibnamefont {Eyert}},
  \bibinfo {author} {\bibfnamefont {M.}~\bibnamefont {Knupfer}}, \bibinfo
  {author} {\bibfnamefont {J.}~\bibnamefont {Fink}}, \ and\ \bibinfo {author}
  {\bibfnamefont {J.~F.}\ \bibnamefont {Armbruster}},\ }\href
  {http://stacks.iop.org/0953-8984/8/i=15/a=007} {\bibfield  {journal}
  {\bibinfo  {journal} {J. Phys.: Condens. Matter}\ }\textbf {\bibinfo {volume}
  {8}},\ \bibinfo {pages} {2557} (\bibinfo {year}
  {1996}{\natexlab{b}})}\BibitemShut {NoStop}%
\bibitem [{\citenamefont {Cudazzo}\ \emph {et~al.}(2011)\citenamefont
  {Cudazzo}, \citenamefont {Gatti}, \citenamefont {Roth}, \citenamefont
  {Mahns}, \citenamefont {Knupfer},\ and\ \citenamefont {Rubio}}]{Cudazzo2011}%
  \BibitemOpen
  \bibfield  {author} {\bibinfo {author} {\bibfnamefont {P.}~\bibnamefont
  {Cudazzo}}, \bibinfo {author} {\bibfnamefont {M.}~\bibnamefont {Gatti}},
  \bibinfo {author} {\bibfnamefont {F.}~\bibnamefont {Roth}}, \bibinfo {author}
  {\bibfnamefont {B.}~\bibnamefont {Mahns}}, \bibinfo {author} {\bibfnamefont
  {M.}~\bibnamefont {Knupfer}}, \ and\ \bibinfo {author} {\bibfnamefont
  {A.}~\bibnamefont {Rubio}},\ }\href {\doibase 10.1103/PhysRevB.84.155118}
  {\bibfield  {journal} {\bibinfo  {journal} {Phys. Rev. B}\ }\textbf {\bibinfo
  {volume} {84}},\ \bibinfo {pages} {155118} (\bibinfo {year}
  {2011})}\BibitemShut {NoStop}%
\bibitem [{\citenamefont {Kresin}\ and\ \citenamefont
  {Kresin}(1994)}]{Kresin1994}%
  \BibitemOpen
  \bibfield  {author} {\bibinfo {author} {\bibfnamefont {V.~V.}\ \bibnamefont
  {Kresin}}\ and\ \bibinfo {author} {\bibfnamefont {V.~Z.}\ \bibnamefont
  {Kresin}},\ }\href {\doibase 10.1103/PhysRevB.49.2715} {\bibfield  {journal}
  {\bibinfo  {journal} {Phys. Rev. B}\ }\textbf {\bibinfo {volume} {49}},\
  \bibinfo {pages} {2715} (\bibinfo {year} {1994})}\BibitemShut {NoStop}%
\bibitem [{\citenamefont {Wilson}\ and\ \citenamefont
  {Yoffe}(1969)}]{Wilson1969}%
  \BibitemOpen
  \bibfield  {author} {\bibinfo {author} {\bibfnamefont {J.~A.}\ \bibnamefont
  {Wilson}}\ and\ \bibinfo {author} {\bibfnamefont {A.~D.}\ \bibnamefont
  {Yoffe}},\ }\href {http://www.informaworld.com/10.1080/00018736900101307}
  {\bibfield  {journal} {\bibinfo  {journal} {Adv. Phys.}\ }\textbf {\bibinfo
  {volume} {18}},\ \bibinfo {pages} {193} (\bibinfo {year} {1969})}\BibitemShut
  {NoStop}%
\bibitem [{\citenamefont {Wilson}\ \emph {et~al.}(1975)\citenamefont {Wilson},
  \citenamefont {Di~Salvo},\ and\ \citenamefont {Mahajan}}]{Wilson1975}%
  \BibitemOpen
  \bibfield  {author} {\bibinfo {author} {\bibfnamefont {J.}~\bibnamefont
  {Wilson}}, \bibinfo {author} {\bibfnamefont {F.}~\bibnamefont {Di~Salvo}}, \
  and\ \bibinfo {author} {\bibfnamefont {S.}~\bibnamefont {Mahajan}},\ }\href
  {\doibase 10.1080/00018737500101391} {\bibfield  {journal} {\bibinfo
  {journal} {Adv. Phys.}\ }\textbf {\bibinfo {volume} {24}},\ \bibinfo {pages}
  {117} (\bibinfo {year} {1975})}\BibitemShut {NoStop}%
\bibitem [{\citenamefont {Friend}\ and\ \citenamefont
  {Yoffe}(1987)}]{Friend1987}%
  \BibitemOpen
  \bibfield  {author} {\bibinfo {author} {\bibfnamefont {R.}~\bibnamefont
  {Friend}}\ and\ \bibinfo {author} {\bibfnamefont {A.}~\bibnamefont {Yoffe}},\
  }\href {\doibase 10.1080/00018738700101951} {\bibfield  {journal} {\bibinfo
  {journal} {Adv. Phys.}\ }\textbf {\bibinfo {volume} {36}},\ \bibinfo {pages}
  {1} (\bibinfo {year} {1987})}\BibitemShut {NoStop}%
\bibitem [{\citenamefont {Moncton}\ \emph {et~al.}(1975)\citenamefont
  {Moncton}, \citenamefont {Axe},\ and\ \citenamefont {{Di
  Salvo}}}]{Moncton1975}%
  \BibitemOpen
  \bibfield  {author} {\bibinfo {author} {\bibfnamefont {D.~E.}\ \bibnamefont
  {Moncton}}, \bibinfo {author} {\bibfnamefont {J.~D.}\ \bibnamefont {Axe}}, \
  and\ \bibinfo {author} {\bibfnamefont {F.~J.}\ \bibnamefont {{Di Salvo}}},\
  }\href {\doibase 10.1103/PhysRevLett.34.734} {\bibfield  {journal} {\bibinfo
  {journal} {Phys. Rev. Lett.}\ }\textbf {\bibinfo {volume} {34}},\ \bibinfo
  {pages} {734} (\bibinfo {year} {1975})}\BibitemShut {NoStop}%
\bibitem [{\citenamefont {van Wezel}\ \emph {et~al.}(2011)\citenamefont {van
  Wezel}, \citenamefont {Schuster}, \citenamefont {K{\"o}nig}, \citenamefont
  {Knupfer}, \citenamefont {van~den Brink}, \citenamefont {Berger},\ and\
  \citenamefont {B{\"u}chner}}]{vanWezel2011a}%
  \BibitemOpen
  \bibfield  {author} {\bibinfo {author} {\bibfnamefont {J.}~\bibnamefont {van
  Wezel}}, \bibinfo {author} {\bibfnamefont {R.}~\bibnamefont {Schuster}},
  \bibinfo {author} {\bibfnamefont {A.}~\bibnamefont {K{\"o}nig}}, \bibinfo
  {author} {\bibfnamefont {M.}~\bibnamefont {Knupfer}}, \bibinfo {author}
  {\bibfnamefont {J.}~\bibnamefont {van~den Brink}}, \bibinfo {author}
  {\bibfnamefont {H.}~\bibnamefont {Berger}}, \ and\ \bibinfo {author}
  {\bibfnamefont {B.}~\bibnamefont {B{\"u}chner}},\ }\href {\doibase
  10.1103/PhysRevLett.107.176404} {\bibfield  {journal} {\bibinfo  {journal}
  {Phys. Rev. Lett.}\ }\textbf {\bibinfo {volume} {107}},\ \bibinfo {pages}
  {176404} (\bibinfo {year} {2011})}\BibitemShut {NoStop}%
\bibitem [{\citenamefont {vom Felde}\ \emph {et~al.}(1987)\citenamefont {vom
  Felde}, \citenamefont {Fink}, \citenamefont {B\"uche}, \citenamefont
  {Scheerer},\ and\ \citenamefont {N\"ucker}}]{Felde1987}%
  \BibitemOpen
  \bibfield  {author} {\bibinfo {author} {\bibfnamefont {A.}~\bibnamefont {vom
  Felde}}, \bibinfo {author} {\bibfnamefont {J.}~\bibnamefont {Fink}}, \bibinfo
  {author} {\bibfnamefont {T.}~\bibnamefont {B\"uche}}, \bibinfo {author}
  {\bibfnamefont {B.}~\bibnamefont {Scheerer}}, \ and\ \bibinfo {author}
  {\bibfnamefont {N.}~\bibnamefont {N\"ucker}},\ }\href
  {http://stacks.iop.org/0295-5075/4/i=9/a=014} {\bibfield  {journal} {\bibinfo
   {journal} {Europhys. Lett.}\ }\textbf {\bibinfo {volume} {4}},\ \bibinfo
  {pages} {1037} (\bibinfo {year} {1987})}\BibitemShut {NoStop}%
\bibitem [{\citenamefont {vom Felde}\ \emph {et~al.}(1989)\citenamefont {vom
  Felde}, \citenamefont {Spr\"osser-Prou},\ and\ \citenamefont
  {Fink}}]{Felde1989}%
  \BibitemOpen
  \bibfield  {author} {\bibinfo {author} {\bibfnamefont {A.}~\bibnamefont {vom
  Felde}}, \bibinfo {author} {\bibfnamefont {J.}~\bibnamefont
  {Spr\"osser-Prou}}, \ and\ \bibinfo {author} {\bibfnamefont {J.}~\bibnamefont
  {Fink}},\ }\href {\doibase 10.1103/PhysRevB.40.10181} {\bibfield  {journal}
  {\bibinfo  {journal} {Phys. Rev. B}\ }\textbf {\bibinfo {volume} {40}},\
  \bibinfo {pages} {10181} (\bibinfo {year} {1989})}\BibitemShut {NoStop}%
\bibitem [{\citenamefont {Manzke}\ \emph {et~al.}(1981)\citenamefont {Manzke},
  \citenamefont {Crecelius}, \citenamefont {Fink},\ and\ \citenamefont
  {Sch{\"o}llhorn}}]{Manzke1981}%
  \BibitemOpen
  \bibfield  {author} {\bibinfo {author} {\bibfnamefont {R.}~\bibnamefont
  {Manzke}}, \bibinfo {author} {\bibfnamefont {G.}~\bibnamefont {Crecelius}},
  \bibinfo {author} {\bibfnamefont {J.}~\bibnamefont {Fink}}, \ and\ \bibinfo
  {author} {\bibfnamefont {R.}~\bibnamefont {Sch{\"o}llhorn}},\ }\href
  {http://www.sciencedirect.com/science/article/B6TVW-46MF9G0-1BT/1/dff54062a8501586a5c30affb2d28f46}
  {\bibfield  {journal} {\bibinfo  {journal} {Solid State Commun.}\ }\textbf
  {\bibinfo {volume} {40}},\ \bibinfo {pages} {103} (\bibinfo {year}
  {1981})}\BibitemShut {NoStop}%
\bibitem [{\citenamefont {Cudazzo}\ \emph {et~al.}(2012)\citenamefont
  {Cudazzo}, \citenamefont {Gatti},\ and\ \citenamefont {Rubio}}]{Cudazzo2012}%
  \BibitemOpen
  \bibfield  {author} {\bibinfo {author} {\bibfnamefont {P.}~\bibnamefont
  {Cudazzo}}, \bibinfo {author} {\bibfnamefont {M.}~\bibnamefont {Gatti}}, \
  and\ \bibinfo {author} {\bibfnamefont {A.}~\bibnamefont {Rubio}},\ }\href
  {\doibase 10.1103/PhysRevB.86.075121} {\bibfield  {journal} {\bibinfo
  {journal} {Phys. Rev. B}\ }\textbf {\bibinfo {volume} {86}},\ \bibinfo
  {pages} {075121} (\bibinfo {year} {2012})}\BibitemShut {NoStop}%
\bibitem [{\citenamefont {Bednorz}\ and\ \citenamefont
  {M{\"u}ller}(1986)}]{Bednorz1986}%
  \BibitemOpen
  \bibfield  {author} {\bibinfo {author} {\bibfnamefont {J.}~\bibnamefont
  {Bednorz}}\ and\ \bibinfo {author} {\bibfnamefont {K.}~\bibnamefont
  {M{\"u}ller}},\ }\href@noop {} {\bibfield  {journal} {\bibinfo  {journal} {Z.
  Phys. B}\ }\textbf {\bibinfo {volume} {64}},\ \bibinfo {pages} {189}
  (\bibinfo {year} {1986})}\BibitemShut {NoStop}%
\bibitem [{\citenamefont {Knupfer}\ \emph {et~al.}(2004)\citenamefont
  {Knupfer}, \citenamefont {Fink}, \citenamefont {Drechsler}, \citenamefont
  {Hayn}, \citenamefont {Malek},\ and\ \citenamefont
  {Moskvin}}]{Knupfer2004_cuprate}%
  \BibitemOpen
  \bibfield  {author} {\bibinfo {author} {\bibfnamefont {M.}~\bibnamefont
  {Knupfer}}, \bibinfo {author} {\bibfnamefont {J.}~\bibnamefont {Fink}},
  \bibinfo {author} {\bibfnamefont {S.-L.}\ \bibnamefont {Drechsler}}, \bibinfo
  {author} {\bibfnamefont {R.}~\bibnamefont {Hayn}}, \bibinfo {author}
  {\bibfnamefont {J.}~\bibnamefont {Malek}}, \ and\ \bibinfo {author}
  {\bibfnamefont {A.}~\bibnamefont {Moskvin}},\ }\href {\doibase
  http://dx.doi.org/10.1016/j.elspec.2004.02.080} {\bibfield  {journal}
  {\bibinfo  {journal} {J. Electron. Spectrosc. Relat. Phenom.}\ }\textbf
  {\bibinfo {volume} {137–140}},\ \bibinfo {pages} {469 } (\bibinfo {year}
  {2004})}\BibitemShut {NoStop}%
\bibitem [{\citenamefont {Uehara}\ \emph {et~al.}(1996)\citenamefont {Uehara},
  \citenamefont {Nagata}, \citenamefont {Akimitsu}, \citenamefont {Takahashi},
  \citenamefont {Mori},\ and\ \citenamefont {Kinoshita}}]{Uehara1996}%
  \BibitemOpen
  \bibfield  {author} {\bibinfo {author} {\bibfnamefont {M.}~\bibnamefont
  {Uehara}}, \bibinfo {author} {\bibfnamefont {T.}~\bibnamefont {Nagata}},
  \bibinfo {author} {\bibfnamefont {J.}~\bibnamefont {Akimitsu}}, \bibinfo
  {author} {\bibfnamefont {H.}~\bibnamefont {Takahashi}}, \bibinfo {author}
  {\bibfnamefont {N.}~\bibnamefont {Mori}}, \ and\ \bibinfo {author}
  {\bibfnamefont {K.}~\bibnamefont {Kinoshita}},\ }\href@noop {} {\bibfield
  {journal} {\bibinfo  {journal} {J. Phys. Soc. Jpn.}\ }\textbf {\bibinfo
  {volume} {65}},\ \bibinfo {pages} {2764} (\bibinfo {year}
  {1996})}\BibitemShut {NoStop}%
\bibitem [{\citenamefont {Nagata}\ \emph {et~al.}(1997)\citenamefont {Nagata},
  \citenamefont {Uehara}, \citenamefont {Goto}, \citenamefont {Komiya},
  \citenamefont {Akimitsu}, \citenamefont {Motoyama}, \citenamefont {Eisaki},
  \citenamefont {Uchida}, \citenamefont {Takahashi}, \citenamefont
  {Nakanishi},\ and\ \citenamefont {M\^ori}}]{Nagata1997}%
  \BibitemOpen
  \bibfield  {author} {\bibinfo {author} {\bibfnamefont {T.}~\bibnamefont
  {Nagata}}, \bibinfo {author} {\bibfnamefont {M.}~\bibnamefont {Uehara}},
  \bibinfo {author} {\bibfnamefont {J.}~\bibnamefont {Goto}}, \bibinfo {author}
  {\bibfnamefont {N.}~\bibnamefont {Komiya}}, \bibinfo {author} {\bibfnamefont
  {J.}~\bibnamefont {Akimitsu}}, \bibinfo {author} {\bibfnamefont
  {N.}~\bibnamefont {Motoyama}}, \bibinfo {author} {\bibfnamefont
  {H.}~\bibnamefont {Eisaki}}, \bibinfo {author} {\bibfnamefont
  {S.}~\bibnamefont {Uchida}}, \bibinfo {author} {\bibfnamefont
  {H.}~\bibnamefont {Takahashi}}, \bibinfo {author} {\bibfnamefont
  {T.}~\bibnamefont {Nakanishi}}, \ and\ \bibinfo {author} {\bibfnamefont
  {N.}~\bibnamefont {M\^ori}},\ }\href {\doibase DOI:
  10.1016/S0921-4534(97)00247-5} {\bibfield  {journal} {\bibinfo  {journal}
  {Physica C: Superconductivity}\ }\textbf {\bibinfo {volume} {282-287}},\
  \bibinfo {pages} {153 } (\bibinfo {year} {1997})}\BibitemShut {NoStop}%
\bibitem [{\citenamefont {Kato}\ \emph {et~al.}(1996)\citenamefont {Kato},
  \citenamefont {Shiota},\ and\ \citenamefont {Koike}}]{Kato1996}%
  \BibitemOpen
  \bibfield  {author} {\bibinfo {author} {\bibfnamefont {M.}~\bibnamefont
  {Kato}}, \bibinfo {author} {\bibfnamefont {K.}~\bibnamefont {Shiota}}, \ and\
  \bibinfo {author} {\bibfnamefont {Y.}~\bibnamefont {Koike}},\ }\href
  {\doibase DOI: 10.1016/0921-4534(95)00802-0} {\bibfield  {journal} {\bibinfo
  {journal} {Physica C: Superconductivity}\ }\textbf {\bibinfo {volume}
  {258}},\ \bibinfo {pages} {284 } (\bibinfo {year} {1996})}\BibitemShut
  {NoStop}%
\bibitem [{\citenamefont {N\"ucker}\ \emph {et~al.}(2000)\citenamefont
  {N\"ucker}, \citenamefont {Merz}, \citenamefont {Kuntscher}, \citenamefont
  {Gerhold}, \citenamefont {Schuppler}, \citenamefont {Neudert}, \citenamefont
  {Golden}, \citenamefont {Fink}, \citenamefont {Schild}, \citenamefont
  {Stadler}, \citenamefont {Chakarian}, \citenamefont {Freeland}, \citenamefont
  {Idzerda}, \citenamefont {Conder}, \citenamefont {Uehara}, \citenamefont
  {Nagata}, \citenamefont {Goto}, \citenamefont {Akimitsu}, \citenamefont
  {Motoyama}, \citenamefont {Eisaki}, \citenamefont {Uchida}, \citenamefont
  {Ammerahl},\ and\ \citenamefont {Revcolevschi}}]{Nuecker2000}%
  \BibitemOpen
  \bibfield  {author} {\bibinfo {author} {\bibfnamefont {N.}~\bibnamefont
  {N\"ucker}}, \bibinfo {author} {\bibfnamefont {M.}~\bibnamefont {Merz}},
  \bibinfo {author} {\bibfnamefont {C.~A.}\ \bibnamefont {Kuntscher}}, \bibinfo
  {author} {\bibfnamefont {S.}~\bibnamefont {Gerhold}}, \bibinfo {author}
  {\bibfnamefont {S.}~\bibnamefont {Schuppler}}, \bibinfo {author}
  {\bibfnamefont {R.}~\bibnamefont {Neudert}}, \bibinfo {author} {\bibfnamefont
  {M.~S.}\ \bibnamefont {Golden}}, \bibinfo {author} {\bibfnamefont
  {J.}~\bibnamefont {Fink}}, \bibinfo {author} {\bibfnamefont {D.}~\bibnamefont
  {Schild}}, \bibinfo {author} {\bibfnamefont {S.}~\bibnamefont {Stadler}},
  \bibinfo {author} {\bibfnamefont {V.}~\bibnamefont {Chakarian}}, \bibinfo
  {author} {\bibfnamefont {J.}~\bibnamefont {Freeland}}, \bibinfo {author}
  {\bibfnamefont {Y.~U.}\ \bibnamefont {Idzerda}}, \bibinfo {author}
  {\bibfnamefont {K.}~\bibnamefont {Conder}}, \bibinfo {author} {\bibfnamefont
  {M.}~\bibnamefont {Uehara}}, \bibinfo {author} {\bibfnamefont
  {T.}~\bibnamefont {Nagata}}, \bibinfo {author} {\bibfnamefont
  {J.}~\bibnamefont {Goto}}, \bibinfo {author} {\bibfnamefont {J.}~\bibnamefont
  {Akimitsu}}, \bibinfo {author} {\bibfnamefont {N.}~\bibnamefont {Motoyama}},
  \bibinfo {author} {\bibfnamefont {H.}~\bibnamefont {Eisaki}}, \bibinfo
  {author} {\bibfnamefont {S.}~\bibnamefont {Uchida}}, \bibinfo {author}
  {\bibfnamefont {U.}~\bibnamefont {Ammerahl}}, \ and\ \bibinfo {author}
  {\bibfnamefont {A.}~\bibnamefont {Revcolevschi}},\ }\href {\doibase
  10.1103/PhysRevB.62.14384} {\bibfield  {journal} {\bibinfo  {journal} {Phys.
  Rev. B}\ }\textbf {\bibinfo {volume} {62}},\ \bibinfo {pages} {14384}
  (\bibinfo {year} {2000})}\BibitemShut {NoStop}%
\bibitem [{\citenamefont {Koitzsch}\ \emph {et~al.}(2010)\citenamefont
  {Koitzsch}, \citenamefont {Inosov}, \citenamefont {Shiozawa}, \citenamefont
  {Zabolotnyy}, \citenamefont {Borisenko}, \citenamefont {Varykhalov},
  \citenamefont {Hess}, \citenamefont {Knupfer}, \citenamefont {Ammerahl},
  \citenamefont {Revcolevschi},\ and\ \citenamefont
  {B\"uchner}}]{Koitzsch2010}%
  \BibitemOpen
  \bibfield  {author} {\bibinfo {author} {\bibfnamefont {A.}~\bibnamefont
  {Koitzsch}}, \bibinfo {author} {\bibfnamefont {D.~S.}\ \bibnamefont
  {Inosov}}, \bibinfo {author} {\bibfnamefont {H.}~\bibnamefont {Shiozawa}},
  \bibinfo {author} {\bibfnamefont {V.~B.}\ \bibnamefont {Zabolotnyy}},
  \bibinfo {author} {\bibfnamefont {S.~V.}\ \bibnamefont {Borisenko}}, \bibinfo
  {author} {\bibfnamefont {A.}~\bibnamefont {Varykhalov}}, \bibinfo {author}
  {\bibfnamefont {C.}~\bibnamefont {Hess}}, \bibinfo {author} {\bibfnamefont
  {M.}~\bibnamefont {Knupfer}}, \bibinfo {author} {\bibfnamefont
  {U.}~\bibnamefont {Ammerahl}}, \bibinfo {author} {\bibfnamefont
  {A.}~\bibnamefont {Revcolevschi}}, \ and\ \bibinfo {author} {\bibfnamefont
  {B.}~\bibnamefont {B\"uchner}},\ }\href@noop {} {\bibfield  {journal}
  {\bibinfo  {journal} {Phys. Rev. B}\ }\textbf {\bibinfo {volume} {81}},\
  \bibinfo {pages} {113110} (\bibinfo {year} {2010})}\BibitemShut {NoStop}%
\bibitem [{\citenamefont {McCarron}\ \emph {et~al.}(1988)\citenamefont
  {McCarron}, \citenamefont {Subramanian}, \citenamefont {Calabrese},\ and\
  \citenamefont {Harlow}}]{McCarron1988}%
  \BibitemOpen
  \bibfield  {author} {\bibinfo {author} {\bibfnamefont {E.}~\bibnamefont
  {McCarron}}, \bibinfo {author} {\bibfnamefont {M.}~\bibnamefont
  {Subramanian}}, \bibinfo {author} {\bibfnamefont {J.}~\bibnamefont
  {Calabrese}}, \ and\ \bibinfo {author} {\bibfnamefont {R.}~\bibnamefont
  {Harlow}},\ }\href@noop {} {\bibfield  {journal} {\bibinfo  {journal} {Mater.
  Res. Bull.}\ }\textbf {\bibinfo {volume} {23}},\ \bibinfo {pages} {1355}
  (\bibinfo {year} {1988})}\BibitemShut {NoStop}%
\bibitem [{\citenamefont {Siegrist}\ \emph {et~al.}(1988)\citenamefont
  {Siegrist}, \citenamefont {Schneemeyer}, \citenamefont {Sunshine},
  \citenamefont {Waszczak},\ and\ \citenamefont {Roth}}]{Siegrist1988}%
  \BibitemOpen
  \bibfield  {author} {\bibinfo {author} {\bibfnamefont {T.}~\bibnamefont
  {Siegrist}}, \bibinfo {author} {\bibfnamefont {L.}~\bibnamefont
  {Schneemeyer}}, \bibinfo {author} {\bibfnamefont {S.}~\bibnamefont
  {Sunshine}}, \bibinfo {author} {\bibfnamefont {J.}~\bibnamefont {Waszczak}},
  \ and\ \bibinfo {author} {\bibfnamefont {R.}~\bibnamefont {Roth}},\
  }\href@noop {} {\bibfield  {journal} {\bibinfo  {journal} {Mater. Res.
  Bull.}\ }\textbf {\bibinfo {volume} {23}},\ \bibinfo {pages} {1429} (\bibinfo
  {year} {1988})}\BibitemShut {NoStop}%
\bibitem [{\citenamefont {Motoyama}\ \emph {et~al.}(1997)\citenamefont
  {Motoyama}, \citenamefont {Osafune}, \citenamefont {Kakeshita}, \citenamefont
  {Eisaki},\ and\ \citenamefont {Uchida}}]{Motoyama1997}%
  \BibitemOpen
  \bibfield  {author} {\bibinfo {author} {\bibfnamefont {N.}~\bibnamefont
  {Motoyama}}, \bibinfo {author} {\bibfnamefont {T.}~\bibnamefont {Osafune}},
  \bibinfo {author} {\bibfnamefont {T.}~\bibnamefont {Kakeshita}}, \bibinfo
  {author} {\bibfnamefont {H.}~\bibnamefont {Eisaki}}, \ and\ \bibinfo {author}
  {\bibfnamefont {S.}~\bibnamefont {Uchida}},\ }\href {\doibase
  10.1103/PhysRevB.55.R3386} {\bibfield  {journal} {\bibinfo  {journal} {Phys.
  Rev. B}\ }\textbf {\bibinfo {volume} {55}},\ \bibinfo {pages} {R3386}
  (\bibinfo {year} {1997})}\BibitemShut {NoStop}%
\bibitem [{\citenamefont {Osafune}\ \emph {et~al.}(1997)\citenamefont
  {Osafune}, \citenamefont {Motoyama}, \citenamefont {Eisaki},\ and\
  \citenamefont {Uchida}}]{Osafune1997}%
  \BibitemOpen
  \bibfield  {author} {\bibinfo {author} {\bibfnamefont {T.}~\bibnamefont
  {Osafune}}, \bibinfo {author} {\bibfnamefont {N.}~\bibnamefont {Motoyama}},
  \bibinfo {author} {\bibfnamefont {H.}~\bibnamefont {Eisaki}}, \ and\ \bibinfo
  {author} {\bibfnamefont {S.}~\bibnamefont {Uchida}},\ }\href {\doibase
  10.1103/PhysRevLett.78.1980} {\bibfield  {journal} {\bibinfo  {journal}
  {Phys. Rev. Lett.}\ }\textbf {\bibinfo {volume} {78}},\ \bibinfo {pages}
  {1980} (\bibinfo {year} {1997})}\BibitemShut {NoStop}%
\bibitem [{\citenamefont {Ruzicka}\ \emph {et~al.}(1998)\citenamefont
  {Ruzicka}, \citenamefont {Degiorgi}, \citenamefont {Ammerahl}, \citenamefont
  {Dhalenne},\ and\ \citenamefont {Revcolevschi}}]{Ruzicka1998}%
  \BibitemOpen
  \bibfield  {author} {\bibinfo {author} {\bibfnamefont {B.}~\bibnamefont
  {Ruzicka}}, \bibinfo {author} {\bibfnamefont {L.}~\bibnamefont {Degiorgi}},
  \bibinfo {author} {\bibfnamefont {U.}~\bibnamefont {Ammerahl}}, \bibinfo
  {author} {\bibfnamefont {G.}~\bibnamefont {Dhalenne}}, \ and\ \bibinfo
  {author} {\bibfnamefont {A.}~\bibnamefont {Revcolevschi}},\ }\href@noop {}
  {\bibfield  {journal} {\bibinfo  {journal} {Eur. Phys. J. B}\ }\textbf
  {\bibinfo {volume} {6}},\ \bibinfo {pages} {301} (\bibinfo {year}
  {1998})}\BibitemShut {NoStop}%
\bibitem [{\citenamefont {Wang}\ \emph {et~al.}(1990)\citenamefont {Wang},
  \citenamefont {Feng},\ and\ \citenamefont {Ritter}}]{Wang1990}%
  \BibitemOpen
  \bibfield  {author} {\bibinfo {author} {\bibfnamefont {Y.-Y.}\ \bibnamefont
  {Wang}}, \bibinfo {author} {\bibfnamefont {G.}~\bibnamefont {Feng}}, \ and\
  \bibinfo {author} {\bibfnamefont {A.~L.}\ \bibnamefont {Ritter}},\ }\href
  {\doibase 10.1103/PhysRevB.42.420} {\bibfield  {journal} {\bibinfo  {journal}
  {Phys. Rev. B}\ }\textbf {\bibinfo {volume} {42}},\ \bibinfo {pages} {420}
  (\bibinfo {year} {1990})}\BibitemShut {NoStop}%
\bibitem [{\citenamefont {N{\"u}cker}\ \emph {et~al.}(1991)\citenamefont
  {N{\"u}cker}, \citenamefont {Eckern}, \citenamefont {Fink},\ and\
  \citenamefont {M{\"u}ller}}]{Nuecker1991}%
  \BibitemOpen
  \bibfield  {author} {\bibinfo {author} {\bibfnamefont {N.}~\bibnamefont
  {N{\"u}cker}}, \bibinfo {author} {\bibfnamefont {U.}~\bibnamefont {Eckern}},
  \bibinfo {author} {\bibfnamefont {J.}~\bibnamefont {Fink}}, \ and\ \bibinfo
  {author} {\bibfnamefont {P.}~\bibnamefont {M{\"u}ller}},\ }\href {\doibase
  10.1103/PhysRevB.44.7155} {\bibfield  {journal} {\bibinfo  {journal} {Phys.
  Rev. B}\ }\textbf {\bibinfo {volume} {44}},\ \bibinfo {pages} {7155}
  (\bibinfo {year} {1991})}\BibitemShut {NoStop}%
\bibitem [{\citenamefont {Knupfer}\ \emph {et~al.}(1994)\citenamefont
  {Knupfer}, \citenamefont {Roth}, \citenamefont {Fink}, \citenamefont
  {Karpinski},\ and\ \citenamefont {Kaldis}}]{Knupfer1994}%
  \BibitemOpen
  \bibfield  {author} {\bibinfo {author} {\bibfnamefont {M.}~\bibnamefont
  {Knupfer}}, \bibinfo {author} {\bibfnamefont {G.}~\bibnamefont {Roth}},
  \bibinfo {author} {\bibfnamefont {J.}~\bibnamefont {Fink}}, \bibinfo {author}
  {\bibfnamefont {J.}~\bibnamefont {Karpinski}}, \ and\ \bibinfo {author}
  {\bibfnamefont {E.}~\bibnamefont {Kaldis}},\ }\href {\doibase DOI:
  10.1016/0921-4534(94)90453-7} {\bibfield  {journal} {\bibinfo  {journal}
  {Physica C: Superconductivity}\ }\textbf {\bibinfo {volume} {230}},\ \bibinfo
  {pages} {121} (\bibinfo {year} {1994})}\BibitemShut {NoStop}%
\bibitem [{\citenamefont {Williams}\ and\ \citenamefont
  {Bloch}(1974)}]{Williams1974}%
  \BibitemOpen
  \bibfield  {author} {\bibinfo {author} {\bibfnamefont {P.~F.}\ \bibnamefont
  {Williams}}\ and\ \bibinfo {author} {\bibfnamefont {A.~N.}\ \bibnamefont
  {Bloch}},\ }\href {\doibase 10.1103/PhysRevB.10.1097} {\bibfield  {journal}
  {\bibinfo  {journal} {Phys. Rev. B}\ }\textbf {\bibinfo {volume} {10}},\
  \bibinfo {pages} {1097} (\bibinfo {year} {1974})}\BibitemShut {NoStop}%
\bibitem [{\citenamefont {Sing}\ \emph {et~al.}(1998)\citenamefont {Sing},
  \citenamefont {Grigoryan}, \citenamefont {Paasch}, \citenamefont {Knupfer},
  \citenamefont {Fink}, \citenamefont {Berger},\ and\ \citenamefont
  {L\'evy}}]{Sing1998}%
  \BibitemOpen
  \bibfield  {author} {\bibinfo {author} {\bibfnamefont {M.}~\bibnamefont
  {Sing}}, \bibinfo {author} {\bibfnamefont {V.~G.}\ \bibnamefont {Grigoryan}},
  \bibinfo {author} {\bibfnamefont {G.}~\bibnamefont {Paasch}}, \bibinfo
  {author} {\bibfnamefont {M.}~\bibnamefont {Knupfer}}, \bibinfo {author}
  {\bibfnamefont {J.}~\bibnamefont {Fink}}, \bibinfo {author} {\bibfnamefont
  {H.}~\bibnamefont {Berger}}, \ and\ \bibinfo {author} {\bibfnamefont
  {F.}~\bibnamefont {L\'evy}},\ }\href {\doibase 10.1103/PhysRevB.57.12768}
  {\bibfield  {journal} {\bibinfo  {journal} {Phys. Rev. B}\ }\textbf {\bibinfo
  {volume} {57}},\ \bibinfo {pages} {12768} (\bibinfo {year}
  {1998})}\BibitemShut {NoStop}%
\bibitem [{\citenamefont {Sing}\ \emph {et~al.}(1999)\citenamefont {Sing},
  \citenamefont {Grigoryan}, \citenamefont {Paasch}, \citenamefont {Knupfer},
  \citenamefont {Fink}, \citenamefont {Levy}, \citenamefont {Berger},
  \citenamefont {Lommel},\ and\ \citenamefont {Assmus}}]{Sing1999}%
  \BibitemOpen
  \bibfield  {author} {\bibinfo {author} {\bibfnamefont {M.}~\bibnamefont
  {Sing}}, \bibinfo {author} {\bibfnamefont {V.}~\bibnamefont {Grigoryan}},
  \bibinfo {author} {\bibfnamefont {G.}~\bibnamefont {Paasch}}, \bibinfo
  {author} {\bibfnamefont {M.}~\bibnamefont {Knupfer}}, \bibinfo {author}
  {\bibfnamefont {J.}~\bibnamefont {Fink}}, \bibinfo {author} {\bibfnamefont
  {F.}~\bibnamefont {Levy}}, \bibinfo {author} {\bibfnamefont {H.}~\bibnamefont
  {Berger}}, \bibinfo {author} {\bibfnamefont {B.}~\bibnamefont {Lommel}}, \
  and\ \bibinfo {author} {\bibfnamefont {W.}~\bibnamefont {Assmus}},\
  }\href@noop {} {\bibfield  {journal} {\bibinfo  {journal} {Synthetic Metals}\
  }\textbf {\bibinfo {volume} {102}},\ \bibinfo {pages} {1591} (\bibinfo {year}
  {1999})}\BibitemShut {NoStop}%
\bibitem [{\citenamefont {Schuster}(2010)}]{Schuster2010}%
  \BibitemOpen
  \bibfield  {author} {\bibinfo {author} {\bibfnamefont {R.}~\bibnamefont
  {Schuster}},\ }\emph {\bibinfo {title} {Electron Energy-Loss Spectroscopy On
  Underdoped Cuprates And Transition-Metal Dichalcogenides}},\ \href@noop {}
  {Ph.D. thesis},\ \bibinfo  {school} {TU Dresden} (\bibinfo {year}
  {2010})\BibitemShut {NoStop}%
\bibitem [{\citenamefont {Abbamonte}\ \emph {et~al.}(2004)\citenamefont
  {Abbamonte}, \citenamefont {Blumberg}, \citenamefont {Rusydi}, \citenamefont
  {Gozar}, \citenamefont {Evans}, \citenamefont {Siegrist}, \citenamefont
  {Venema}, \citenamefont {Eisaki}, \citenamefont {Isaacs},\ and\ \citenamefont
  {Sawatzky}}]{Abbamonte2004}%
  \BibitemOpen
  \bibfield  {author} {\bibinfo {author} {\bibfnamefont {P.}~\bibnamefont
  {Abbamonte}}, \bibinfo {author} {\bibfnamefont {G.}~\bibnamefont {Blumberg}},
  \bibinfo {author} {\bibfnamefont {A.}~\bibnamefont {Rusydi}}, \bibinfo
  {author} {\bibfnamefont {A.}~\bibnamefont {Gozar}}, \bibinfo {author}
  {\bibfnamefont {P.}~\bibnamefont {Evans}}, \bibinfo {author} {\bibfnamefont
  {T.}~\bibnamefont {Siegrist}}, \bibinfo {author} {\bibfnamefont
  {L.}~\bibnamefont {Venema}}, \bibinfo {author} {\bibfnamefont
  {H.}~\bibnamefont {Eisaki}}, \bibinfo {author} {\bibfnamefont
  {E.}~\bibnamefont {Isaacs}}, \ and\ \bibinfo {author} {\bibfnamefont
  {G.}~\bibnamefont {Sawatzky}},\ }\href@noop {} {\bibfield  {journal}
  {\bibinfo  {journal} {Nature}\ }\textbf {\bibinfo {volume} {431}},\ \bibinfo
  {pages} {1078} (\bibinfo {year} {2004})}\BibitemShut {NoStop}%
\bibitem [{\citenamefont {Carr}\ and\ \citenamefont
  {Tsvelik}(2002)}]{Carr2002}%
  \BibitemOpen
  \bibfield  {author} {\bibinfo {author} {\bibfnamefont {S.~T.}\ \bibnamefont
  {Carr}}\ and\ \bibinfo {author} {\bibfnamefont {A.~M.}\ \bibnamefont
  {Tsvelik}},\ }\href {\doibase 10.1103/PhysRevB.65.195121} {\bibfield
  {journal} {\bibinfo  {journal} {Phys. Rev. B}\ }\textbf {\bibinfo {volume}
  {65}},\ \bibinfo {pages} {195121} (\bibinfo {year} {2002})}\BibitemShut
  {NoStop}%
\bibitem [{\citenamefont {White}\ \emph {et~al.}(2002)\citenamefont {White},
  \citenamefont {Affleck},\ and\ \citenamefont {Scalapino}}]{Friedel2002}%
  \BibitemOpen
  \bibfield  {author} {\bibinfo {author} {\bibfnamefont {S.~R.}\ \bibnamefont
  {White}}, \bibinfo {author} {\bibfnamefont {I.}~\bibnamefont {Affleck}}, \
  and\ \bibinfo {author} {\bibfnamefont {D.~J.}\ \bibnamefont {Scalapino}},\
  }\href {\doibase 10.1103/PhysRevB.65.165122} {\bibfield  {journal} {\bibinfo
  {journal} {Phys. Rev. B}\ }\textbf {\bibinfo {volume} {65}},\ \bibinfo
  {pages} {165122} (\bibinfo {year} {2002})}\BibitemShut {NoStop}%
\bibitem [{\citenamefont {Hess}\ \emph {et~al.}(2004)\citenamefont {Hess},
  \citenamefont {ElHaes}, \citenamefont {B\"uchner}, \citenamefont {Ammerahl},
  \citenamefont {H\"ucker},\ and\ \citenamefont {Revcolevschi}}]{Hess2004}%
  \BibitemOpen
  \bibfield  {author} {\bibinfo {author} {\bibfnamefont {C.}~\bibnamefont
  {Hess}}, \bibinfo {author} {\bibfnamefont {H.}~\bibnamefont {ElHaes}},
  \bibinfo {author} {\bibfnamefont {B.}~\bibnamefont {B\"uchner}}, \bibinfo
  {author} {\bibfnamefont {U.}~\bibnamefont {Ammerahl}}, \bibinfo {author}
  {\bibfnamefont {M.}~\bibnamefont {H\"ucker}}, \ and\ \bibinfo {author}
  {\bibfnamefont {A.}~\bibnamefont {Revcolevschi}},\ }\href {\doibase
  10.1103/PhysRevLett.93.027005} {\bibfield  {journal} {\bibinfo  {journal}
  {Phys. Rev. Lett.}\ }\textbf {\bibinfo {volume} {93}},\ \bibinfo {pages}
  {027005} (\bibinfo {year} {2004})}\BibitemShut {NoStop}%
\end{thebibliography}
\end{document}